\documentclass[11pt,letter]{article}
\pdfoutput=1
\usepackage{jheppub}
\usepackage{amsmath,amssymb}
\usepackage[dvipsnames]{xcolor}
\usepackage{xspace}
\usepackage{ifdraft}
\usepackage{epstopdf}
\usepackage{slashed}
\usepackage{anyfontsize}
\usepackage{tocloft}

\usepackage{diagbox}
\usepackage{colortbl}
\usepackage{bbold}
\usepackage{tabularx}
\usepackage{graphicx}
\usepackage{stackengine}
\usepackage{subcaption}
\usepackage{cleveref}
\usepackage{orcidlink}
\usepackage[normalem]{ulem}

\usepackage{tikz-feynman} 
\usepackage{xcolor}
\hypersetup{
    colorlinks,
    linkcolor={nhpRed},
    citecolor={nhpBlue},
    urlcolor={nhpBlue}
}

\definecolor{nhpRed}{RGB}{161,0,0}
\definecolor{nhp4}{RGB}{203, 4, 31}
\definecolor{nhp3}{RGB}{244,99,30}
\definecolor{nhp2}{RGB}{255,159,0}
\definecolor{nhp1}{RGB}{48,152,152}
\definecolor{nhpBlue}{RGB}{0,100,144}
\definecolor{cutred}{RGB}{219,56,49}
\definecolor{hgreen}{RGB}{25,176,146}
\definecolor{hgreen1}{RGB}{175,230,175}
\definecolor{hblue}{RGB}{52,152,219}
\definecolor{hbluedark}{RGB}{36, 106, 160}
\definecolor{hblue1}{RGB}{255,255,166}
\definecolor{hred}{RGB}{216,83,117}
\definecolor{hreddark}{RGB}{151, 58, 81}
\definecolor{hred1}{RGB}{255,155,155}
\definecolor{cutred}{RGB}{219,56,49}
\definecolor{hgrey4}{RGB}{75,75,75}
\definecolor{hgrey5}{RGB}{50,50,50}
\definecolor{hgrey3}{RGB}{100,100,100}
\definecolor{hgrey}{RGB}{125,125,125}
\definecolor{hgrey2}{RGB}{125,125,125}
\definecolor{hgrey1}{RGB}{150,150,150}
\definecolor{hgrey0}{RGB}{175,175,175}
\definecolor{darkgreen}{RGB}{59,126,108}
\definecolor{nu-80}{RGB}{104, 076, 150}
\definecolor{nu-40}{RGB}{164, 149, 195}
\colorlet{ucp-color}{nhpBlue}
\colorlet{cut-color}{nhp4}
\colorlet{jacobi-color}{nhp4}
\colorlet{refl-color}{nhp4}

\usepackage[framemethod=default]{mdframed}
\newmdenv[skipabove=7pt,
skipbelow=7pt,
rightline=false,
leftline=false,
topline=false,
bottomline=false,
backgroundcolor=gray!15,
linecolor=gray,
innerleftmargin=5pt,
innerrightmargin=5pt,
innertopmargin=5pt,
innerbottommargin=5pt,
leftmargin=0cm,
rightmargin=0cm,
linewidth=4pt]{eBox}

\usetikzlibrary{calc}
\tikzset{Bfield/.style={decorate, double}}

\tikzfeynmanset{
  fermion2/.style={
    /tikz/postaction={
      /tikz/decoration={
        markings,
        mark=at position 0.7 with {
          \arrow{>[length=6pt, width=7pt]};
        },
      },
      /tikz/decorate=true,
    },
  },
}

\tikzset{every picture/.style={baseline={([yshift=-.7ex]current bounding box.center)}}}

\newcommand{\blob}[2]{\vertex[dot,scale=2.4] (#1) at (#2){};\vertex[dot,scale=1.8,hgrey0] at (#2){};}
\newcommand{\ver}[2]{\coordinate (#1) at (#2){};}

\newcommand{\KTDBContrib}[4]{ {
    \begin{tikzpicture}
      \pgfmathsetmacro{\cutlength}{0.8}
\begin{feynman}
\vertex (a1) at (-1.5,1){#4};
\vertex (a2) at (1.5,-1){#2};
\vertex (a3) at (1.5,1){#3};
\vertex (a4) at (-1.5,-1){#1};
\vertex (mid1) at (0,-.5);
\vertex (mid2) at (0,.5);
\vertex (mid3) at (1,.5) ;
\vertex (mid4) at (1,-.5) ;
\vertex (mid5) at (-1,.5);
\vertex (mid6) at (-1,-.5) ;
\vertex (m1) at ($(mid5)!0.5!(mid6)$);
\vertex (m2) at ($(mid1)!0.5!(mid6)$);
\vertex (m3) at ($(mid5)!0.5!(mid2)$);
\vertex (clo) at ($(m2)!0.5!(m3)$) {};
\vertex (c1) at ($(clo)+(90:\cutlength)$);
\vertex (c2) at ($(clo)+(180:\cutlength)$);
\vertex (c3) at ($(clo)+(270:\cutlength)$);
\vertex (m4) at ($(mid3)!0.5!(mid4)$);
\vertex (m5) at ($(mid3)!0.5!(mid2)$);
\vertex (m6) at ($(mid4)!0.5!(mid1)$);
\vertex (cro) at ($(m5)!0.5!(m6)$) {};
\vertex (c4) at ($(cro)+(90:\cutlength)$);
\vertex (c5) at ($(cro)+(0:\cutlength)$);
\vertex (c6) at ($(cro)+(270:\cutlength)$);

\diagram{
(a4) --[ultra thick,](mid6),
(a3) --[ultra thick,](mid3),
(a2) --[ultra thick,](mid4),
(a1) --[ultra thick,](mid5),
(mid1) --[ultra thick,ucp-color](mid2),
(mid3) --[ultra thick,](mid2),
(mid1) --[ultra thick,](mid6),
(mid1) --[ultra thick,](mid4),
(mid5) --[ultra thick,](mid2),
(mid5) --[ultra thick,](mid6),
(mid3) --[ultra thick,](mid4),
(clo)--[ultra thick,dashed,cut-color](c1),
(clo)--[ultra thick,dashed,cut-color](c2),
(clo)--[ultra thick,dashed,cut-color](c3),
(cro)--[ultra thick,dashed,cut-color](c4),
(cro)--[ultra thick,dashed,cut-color](c5),
(cro)--[ultra thick,dashed,cut-color](c6)
};
\end{feynman}
\end{tikzpicture}
}
}

\newcommand{\doubleBoxObstructionB}{ {
\begin{tikzpicture}
\begin{feynman}
\vertex (a1) at (-1.5,1){};
\vertex (a2) at (1.5,-1){};
\vertex (a3) at (1.5,1){};
\vertex (a4) at (-1.5,-1){};
\vertex (mid1) at (0,-.5);
\vertex (mid2) at (0,.5);
\vertex (mid3) at (1,.5) ;
\vertex (mid4) at (1,-.5) ;
\vertex (mid5) at (-1,.5);
\vertex (mid6) at (-1,-.5) ;
\diagram{
(mid6)--[ultra thick,fermion2](a4),
(mid3) --[ultra thick,fermion2](a3),
(mid4)--[ultra thick,fermion2](a2),
(mid5) --[ultra thick,fermion2](a1),
(mid1) --[ultra thick,fermion2](mid2),
(mid2) --[ultra thick,fermion2](mid3),
(mid1) --[ultra thick,fermion2](mid6),
(mid4) --[ultra thick,fermion2](mid1),
(mid2) --[ultra thick,fermion2](mid5),
(mid5) --[ultra thick,fermion2](mid6),
(mid3) --[ultra thick,fermion2](mid4),
};
\vertex [dot,scale=1.8](mid2) at  (-1,-.5){};
\vertex [dot,scale=1.3,white](mid2) at  (-1,-.5){};
\end{feynman}
\end{tikzpicture}
}
}

\newcommand{\doubleBoxObstructionA}{ {
\begin{tikzpicture}
\begin{feynman}
\vertex (a1) at (-1.5,1){};
\vertex (a2) at (1.5,-1){};
\vertex (a3) at (1.5,1){};
\vertex (a4) at (-1.5,-1){};
\vertex (mid1) at (0,-.5);
\vertex (mid2) at (0,.5);
\vertex (mid3) at (1,.5) ;
\vertex (mid4) at (1,-.5) ;
\vertex (mid5) at (-1,.5);
\vertex (mid6) at (-1,-.5) ;
\diagram{
(mid6)--[ultra thick,fermion2](a4),
(mid3) --[ultra thick,fermion2](a3),
(mid4)--[ultra thick,fermion2](a2),
(mid5) --[ultra thick,fermion2](a1),
(mid1) --[ultra thick,fermion2](mid2),
(mid2) --[ultra thick,fermion2](mid3),
(mid1) --[ultra thick,fermion2](mid6),
(mid4) --[ultra thick,fermion2](mid1),
(mid5) --[ultra thick,fermion2](mid2),
(mid6) --[ultra thick,fermion2](mid5),
(mid3) --[ultra thick,fermion2](mid4),
};
\vertex [dot,scale=1.8](mid2) at (0,.5){};
\vertex [dot,scale=1.3,white](mid2) at (0,.5){};
\end{feynman}
\end{tikzpicture}
}
}

\newcommand{\doubleBoxCut}{ {
\begin{tikzpicture}
\begin{feynman}
\vertex (a1) at (-1.5,1){};
\vertex (a2) at (1.5,-1){};
\vertex (a3) at (1.5,1){};
\vertex (a4) at (-1.5,-1){};
\vertex (mid1) at (0,-.5);
\vertex (mid2) at (0,.5);
\vertex (mid3) at (1,.5) ;
\vertex (mid4) at (1,-.5) ;
\vertex (mid5) at (-1,.5);
\vertex (mid6) at (-1,-.5) ;
\diagram{
(a4) --[ultra thick,](mid6),
(a3) --[ultra thick,](mid3),
(a2) --[ultra thick,](mid4),
(a1) --[ultra thick,](mid5),
(mid1) --[ultra thick,](mid2),
(mid3) --[ultra thick,](mid2),
(mid1) --[ultra thick,](mid6),
(mid1) --[ultra thick,](mid4),
(mid5) --[ultra thick,](mid2),
(mid5) --[ultra thick,](mid6),
(mid3) --[ultra thick,](mid4),
};
\end{feynman}
\end{tikzpicture}
}
}

\newcommand{\NMCutA}{ {
\begin{tikzpicture}
\begin{feynman}
\vertex (a1) at (-1.5,1){};
\vertex (a2) at (1.5,-.5){};
\vertex (a3) at (1.5,.5){};
\vertex (a4) at (-1.5,-1){};
\vertex (mid1) at (0,-.5);
\vertex (mid2) at (0,.5);
\vertex (mid3) at (.75,0);
\vertex (mid5) at (-1,.5);
\vertex (mid6) at (-1,-.5);
\vertex (mid7) at (-.5,0) {};
\diagram{
(a4) --[ultra thick,](mid6),
(a3) --[ultra thick,](mid3),
(a2) --[ultra thick,](mid3),
(a1) --[ultra thick,](mid5),
(mid1) --[ultra thick,](mid6),
(mid3) --[ultra thick,](mid2),
(mid1) --[ultra thick,](mid3),
(mid5) --[ultra thick,](mid7),
(mid7) --[ultra thick,](mid1),
(mid2) --[ultra thick,](mid6),
(mid5) --[ultra thick,](mid2),
};
\vertex [dot,scale=2.4](midC) at (.75,0){};
\vertex [dot,scale=1.8,hgrey0](midC) at (.75,0){};
\end{feynman}
\end{tikzpicture}
}
}

\newcommand{\NMCutB}{ {
\begin{tikzpicture}
\begin{feynman}
\vertex (a1) at (-1.5,1){};
\vertex (a2) at (1.5,-.5){};
\vertex (a3) at (1.5,.5){};
\vertex (a4) at (-1.5,-1){};
\vertex (mid1) at (0,-.5);
\vertex (mid2) at (0,.5);
\vertex (mid3) at (.75,0);
\vertex (mid5) at (-1,.5);
\vertex (mid6) at (-1,-.5);
\diagram{
(a4) --[ultra thick,](mid6),
(a3) --[ultra thick,](mid3),
(a2) --[ultra thick,](mid3),
(a1) --[ultra thick,](mid5),
(mid5) --[ultra thick,](mid6),
(mid3) --[ultra thick,](mid2),
(mid1) --[ultra thick,](mid3),
(mid5) --[ultra thick,](mid2),
(mid1) --[ultra thick,](mid6),
(mid1) --[ultra thick,](mid2),
};
\vertex [dot,scale=2.4](midC) at (.75,0){};
\vertex [dot,scale=1.8,hgrey0](midC) at (.75,0){};
\end{feynman}
\end{tikzpicture}
}
}

\newcommand{\NMCutC}{ {
\begin{tikzpicture}
\begin{feynman}
\vertex (a1) at (-1.25,.85){};
\vertex (a2) at (1.25,-.85){};
\vertex (a3) at (1.25,.85){};
\vertex (a4) at (-1.25,-.85){};
\vertex (mid1) at (0,0);
\vertex (mid3) at (.75,.4) ;
\vertex (mid4) at (.75,-.4) ;
\vertex (mid5) at (-.75,.4);
\vertex (mid6) at (-.75,-.4) ;
\diagram{
(a4) --[ultra thick,](mid6),
(a3) --[ultra thick,](mid3),
(a2) --[ultra thick,](mid4),
(a1) --[ultra thick,](mid5),
(mid3) --[ultra thick,](mid1),
(mid1) --[ultra thick,](mid6),
(mid1) --[ultra thick,](mid4),
(mid5) --[ultra thick,](mid1),
(mid5) --[ultra thick,](mid6),
(mid3) --[ultra thick,](mid4),
};
\vertex [dot,scale=2.4](midC) at (0,0){};
\vertex [dot,scale=1.8,hgrey0](midC) at (0,0){};
\end{feynman}
\end{tikzpicture}
}
}

\newcommand{\NMCutD}{ {
\begin{tikzpicture}
\begin{feynman}
\vertex (a1) at (-1.5,1){};
\vertex (a2) at (.5,1){};
\vertex (a3) at (1.5,0){};
\vertex (a4) at (-1.5,-1){};
\vertex (mid1) at (0,-.5);
\vertex (mid2) at (0,.5);
\vertex (mid3) at (.75,0);
\vertex (mid5) at (-1,.5);
\vertex (mid6) at (-1,-.5);
\vertex (mid7) at (-.5,0) {};
\diagram{
(a4) --[ultra thick,](mid6),
(a3) --[ultra thick,](mid3),
(a2) --[ultra thick,](mid2),
(a1) --[ultra thick,](mid5),
(mid1) --[ultra thick,](mid6),
(mid3) --[ultra thick,](mid2),
(mid1) --[ultra thick,](mid3),
(mid5) --[ultra thick,](mid7),
(mid7) --[ultra thick,](mid1),
(mid2) --[ultra thick,](mid6),
(mid5) --[ultra thick,](mid2),
};
\vertex [dot,scale=2.4](midC) at (0,.5){};
\vertex [dot,scale=1.8,hgrey0](midC) at (0,.5){};
\end{feynman}
\end{tikzpicture}
}
}

\newcommand{\NMCutE}{ {
\begin{tikzpicture}
\begin{feynman}
\vertex (a1) at (-1.3,.7){};
\vertex (a2) at (.7,.85){};
\vertex (a3) at (1.5,0){};
\vertex (a4) at(-1.3,-.7){};
\vertex (mid1) at (.05,-.6);
\vertex (mid2) at (.05,.6);
\vertex (mid3) at (.6,0);
\vertex (mid5) at (-.65,.35);
\vertex (mid6) at (-.65,-.35);
\diagram{
(a4) --[ultra thick,](mid6),
(a3) --[ultra thick,](mid3),
(a2) --[ultra thick,](mid2),
(a1) --[ultra thick,](mid5),
(mid5) --[ultra thick,](mid6),
(mid3) --[ultra thick,](mid2),
(mid1) --[ultra thick,](mid3),
(mid5) --[ultra thick,](mid2),
(mid1) --[ultra thick,](mid6),
(mid1) --[ultra thick,](mid2),
};
\vertex [dot,scale=2.4](midC) at (.05,.6){};
\vertex [dot,scale=1.8,hgrey0](midC) at (.05,.6){};
\end{feynman}
\end{tikzpicture}
}
}

\newcommand{\NMCutF}{ {
\begin{tikzpicture}
\begin{feynman}
\vertex (a2) at (.7,.85){};
\vertex (a3) at (.7,-.85){};
\vertex (a4) at (1.2,-.5){};
\vertex (a1) at (1.2,.5){};
\vertex (mid1) at (.05,-.6);
\vertex (mid2) at (.05,.6);
\vertex (mid3) at (.6,0);
\vertex (mid5) at (-.65,.35);
\vertex (mid6) at (-.65,-.35);
\diagram{
(a4) --[ultra thick,](mid3),
(a3) --[ultra thick,](mid1),
(a2) --[ultra thick,](mid2),
(a1) --[ultra thick,](mid3),
(mid6) --[ultra thick,half left](mid5),
(mid6) --[ultra thick,half right](mid5),
(mid3) --[ultra thick,](mid2),
(mid1) --[ultra thick,](mid3),
(mid5) --[ultra thick,](mid2),
(mid1) --[ultra thick,](mid6),
};
\vertex [dot,scale=2.4](midC) at (.6,0){};
\vertex [dot,scale=1.8,hgrey0](midC) at (.6,0){};
\end{feynman}
\end{tikzpicture}
}
}
\newcommand{\NMCutG}{ {
\begin{tikzpicture}
\begin{feynman}
\vertex (a1) at (-.45,1.15){};
\vertex (a2) at (.7,.85){};
\vertex (a3) at (.7,-.85){};
\vertex (a4) at (1.2,0){};
\vertex (mid1) at (.05,-.6);
\vertex (mid2) at (.05,.6);
\vertex (mid3) at (.6,0);
\vertex (mid5) at (-.65,.35);
\vertex (mid6) at (-.65,-.35);
\diagram{
(a4) --[ultra thick,](mid3),
(a3) --[ultra thick,](mid1),
(a2) --[ultra thick,](mid2),
(a1) --[ultra thick,](mid2),
(mid6) --[ultra thick,half left](mid5),
(mid6) --[ultra thick,half right](mid5),
(mid3) --[ultra thick,](mid2),
(mid1) --[ultra thick,](mid3),
(mid5) --[ultra thick,](mid2),
(mid1) --[ultra thick,](mid6),
};
\vertex [dot,scale=2.4](midC) at (.05,.6){};
\vertex [dot,scale=1.8,hgrey0](midC) at (.05,.6){};
\end{feynman}
\end{tikzpicture}
}
}

\newcommand{\NMCutH}{ {
\begin{tikzpicture}
\begin{feynman}
\vertex (a1) at (-1.3,.7){};
\vertex (a2) at (-1.3,.1){};
\vertex (a3) at (1.5,0){};
\vertex (a4) at (-1.3,-.7){};
\vertex (mid1) at (.05,-.6);
\vertex (mid2) at (.05,.6);
\vertex (mid3) at (.6,0);
\vertex (mid5) at (-.65,.35);
\vertex (mid6) at (-.65,-.35);
\diagram{
(a4) --[ultra thick,](mid6),
(a3) --[ultra thick,](mid3),
(a2) --[ultra thick,](mid5),
(a1) --[ultra thick,](mid5),
(mid5) --[ultra thick,](mid6),
(mid3) --[ultra thick,](mid2),
(mid1) --[ultra thick,](mid3),
(mid5) --[ultra thick,](mid2),
(mid1) --[ultra thick,](mid6),
(mid1) --[ultra thick,](mid2),
};
\vertex [dot,scale=2.4](midC) at (-.65,.35){};
\vertex [dot,scale=1.8,hgrey0](midC) at (-.65,.35){};
\end{feynman}
\end{tikzpicture}
}
}

\newcommand{\pentaTriangleCut}{ {
\begin{tikzpicture}
\begin{feynman}
\vertex (a1) at (-.5,1){};
\vertex (a2) at (1.5,-1){};
\vertex (a3) at (1.5,1){};
\vertex (a4) at (-.5,-1){};
\vertex (mid1) at (0,-.5);
\vertex (mid2) at (0,.5);
\vertex (mid3) at (1,.5);
\vertex (mid4) at (1,-.5);
\vertex (mid5) at (1,0);
\vertex (mid6) at (.5,-.5);
\diagram{
(a4) --[ultra thick,](mid1),
(a3) --[ultra thick,](mid3),
(a2) --[ultra thick,](mid4),
(a1) --[ultra thick,](mid2),
(mid5) --[ultra thick,](mid6),
(mid2) --[ultra thick,](mid1),
(mid3) --[ultra thick,](mid2),
(mid1) --[ultra thick,](mid4),
(mid3) --[ultra thick,](mid4),
};
\end{feynman}
\end{tikzpicture}
}
}

\newcommand{\sumoCut}{ {
\begin{tikzpicture}
\begin{feynman}
\vertex (a1) at (1.1*1.1,0){};
\vertex (a2) at (.6*1.1,-1*1.1){};
\vertex (a3) at (-.6*1.1,-1*1.1){};
\vertex (a4) at (-1.1*1.1,0){};
\vertex (mid1) at (.6*1.1,0);
\vertex (mid2) at (.3*1.1,-.5*1.1);
\vertex (mid3) at (-.3*1.1,-.5*1.1);
\vertex (mid4) at (-.6*1.1,0);
\vertex (mid5) at (-.3*1.1,.5*1.1);
\vertex (mid6) at (.3*1.1,.5*1.1);
\diagram{
(a1) --[ultra thick,](mid1),
(a2) --[ultra thick,](mid2),
(a3) --[ultra thick,](mid3),
(a4) --[ultra thick,](mid4),
(mid1) --[ultra thick,](mid2),
(mid2) --[ultra thick,](mid3),
(mid3) --[ultra thick,](mid4),
(mid4) --[ultra thick,](mid5),
(mid1) --[ultra thick,](mid6),
(mid6) --[ultra thick,half right](mid5),
(mid6) --[ultra thick,half left](mid5),
};
\end{feynman}
\end{tikzpicture}
}
}

\newcommand{\MCutC}{ {
\begin{tikzpicture}
\begin{feynman}
\vertex (a2) at (.7,.85){};
\vertex (a3) at (.7,-.85){};
\vertex (a4) at (1.8,-.5){};
\vertex (a1) at (1.8,.5){};
\vertex (mid1) at (.05,-.6);
\vertex (mid2) at (.05,.6);
\vertex (mid3) at (.6,0);
\vertex (mid4) at (1.2,0);
\vertex (mid5) at (-.65,.35);
\vertex (mid6) at (-.65,-.35);
\diagram{
(a4) --[ultra thick,](mid4),
(a3) --[ultra thick,](mid1),
(a2) --[ultra thick,](mid2),
(a1) --[ultra thick,](mid4),
(mid3) --[ultra thick,](mid4),
(mid6) --[ultra thick,half left](mid5),
(mid6) --[ultra thick,half right](mid5),
(mid3) --[ultra thick,](mid2),
(mid1) --[ultra thick,](mid3),
(mid5) --[ultra thick,](mid2),
(mid1) --[ultra thick,](mid6),
};
\end{feynman}
\end{tikzpicture}
}
}

\newcommand{\MCutD}{ {
\begin{tikzpicture}
\begin{feynman}
\vertex (a2) at (-1.3,-.7){};
\vertex (a3) at (.7,-.85){};
\vertex (a4) at (1.8,-.5){};
\vertex (a1) at (1.8,.5){};
\vertex (mid1) at (.05,-.6);
\vertex (mid2) at (.05,.6);
\vertex (mid3) at (.6,0);
\vertex (mid5) at (-.65,.35);
\vertex (mid6) at (-.65,-.35);
\vertex (mid4) at (1.2,0);
\diagram{
(a4) --[ultra thick,](mid4),
(a3) --[ultra thick,](mid1),
(a2) --[ultra thick,](mid6),
(a1) --[ultra thick,](mid4),
(mid3) --[ultra thick,](mid4),
(mid2) --[ultra thick,half left](mid5),
(mid2) --[ultra thick,half right](mid5),
(mid3) --[ultra thick,](mid2),
(mid1) --[ultra thick,](mid3),
(mid5) --[ultra thick,](mid6),
(mid1) --[ultra thick,](mid6),
};
\end{feynman}
\end{tikzpicture}
}
}

\newcommand{\MCutE}{ {
\begin{tikzpicture}
\begin{feynman}
\vertex (a1) at (-1.5,1){};
\vertex (a2) at (1.8,-.5){};
\vertex (a3) at (1.8,.5){};
\vertex (a4) at (-1.5,-1){};
\vertex (mid1) at (0,-.5);
\vertex (mid2) at (0,.5);
\vertex (mid3) at (.6,0);
\vertex (mid4) at (1.2,0);
\vertex (mid5) at (-1,.5);
\vertex (mid6) at (-1,-.5);
\vertex (mid7) at (-.5,0) {};
\diagram{
(a4) --[ultra thick,](mid6),
(a3) --[ultra thick,](mid4),
(a2) --[ultra thick,](mid4),
(a1) --[ultra thick,](mid5),
(mid1) --[ultra thick,](mid6),
(mid3) --[ultra thick,](mid2),
(mid1) --[ultra thick,](mid3),
(mid5) --[ultra thick,](mid7),
(mid7) --[ultra thick,](mid1),
(mid2) --[ultra thick,](mid6),
(mid5) --[ultra thick,](mid2),
(mid3) --[ultra thick,](mid4),
};
\end{feynman}
\end{tikzpicture}
}
}

\newcommand{\MCutF}{ {
\begin{tikzpicture}
\begin{feynman}
\vertex (a1) at (-1.8,.5){};
\vertex (a2) at (1.8,-.5){};
\vertex (a3) at (1.8,.5){};
\vertex (a4) at (-1.8,-.5){};
\vertex (mid1) at (0,-.5);
\vertex (mid2) at (0,.5);
\vertex (mid3) at (.6,0);
\vertex (mid4) at (1.2,0);
\vertex (mid5) at (-1.2,0);
\vertex (mid6) at (-.6,0);
\diagram{
(a4) --[ultra thick,](mid5),
(a3) --[ultra thick,](mid4),
(a2) --[ultra thick,](mid4),
(a1) --[ultra thick,](mid5),
(mid5) --[ultra thick,](mid6),
(mid1) --[ultra thick,](mid6),
(mid2) --[ultra thick,](mid6),
(mid2) --[ultra thick,](mid1),
(mid2) --[ultra thick,](mid3),
(mid1) --[ultra thick,](mid3),
(mid4) --[ultra thick,](mid3),
};
\end{feynman}
\end{tikzpicture}
}
}

\newcommand{\MCutG}{ {
\begin{tikzpicture}
\begin{feynman}
\vertex (a1) at (-1.5,.85){};
\vertex (a2) at (2,-.5){};
\vertex (a3) at (2,.5){};
\vertex (a4) at (-1.5,-.85){};
\vertex (mid1) at (-.25,0);
\vertex (mid2) at (.25,0);
\vertex (mid3) at (1,.0) ;
\vertex (mid4) at (1.5,.0) ;
\vertex (mid5) at (-1,.4);
\vertex (mid6) at (-1,-.4) ;
\diagram{
(a4) --[ultra thick,](mid6),
(a3) --[ultra thick,](mid4),
(a2) --[ultra thick,](mid4),
(a1) --[ultra thick,](mid5),
(mid1) --[ultra thick,](mid2),
(mid3) --[ultra thick,half right](mid2),
(mid3) --[ultra thick,half left](mid2),
(mid1) --[ultra thick,](mid6),
(mid5) --[ultra thick,](mid1),
(mid5) --[ultra thick,](mid6),
(mid3) --[ultra thick,](mid4),
};
\end{feynman}
\end{tikzpicture}
}
}

\newcommand{\MCutH}{ {
\begin{tikzpicture}
\begin{feynman}
\vertex (a1) at (-2,-.5){};
\vertex (a4) at (-2,.5){};
\vertex (a2) at (2,-.5){};
\vertex (a3) at (2,.5){};
\vertex (mid1) at (-.25,0);
\vertex (mid2) at (.25,0);
\vertex (mid3) at (1,.0) ;
\vertex (mid4) at (1.5,.0) ;
\vertex (mid5) at (-1,.0) ;
\vertex (mid6) at (-1.5,.0) ;
\diagram{
(a4) --[ultra thick,](mid6),
(a3) --[ultra thick,](mid4),
(a2) --[ultra thick,](mid4),
(a1) --[ultra thick,](mid6),
(mid1) --[ultra thick,](mid2),
(mid3) --[ultra thick,half right](mid2),
(mid3) --[ultra thick,half left](mid2),
(mid5) --[ultra thick,half right](mid1),
(mid5) --[ultra thick,half left](mid1),
(mid5) --[ultra thick,](mid6),
(mid3) --[ultra thick,](mid4),
};
\end{feynman}
\end{tikzpicture}
}
}

\newcommand{\MCutI}{ {
\begin{tikzpicture}
\begin{feynman}
\vertex (a1) at (-1.5,1){};
\vertex (a2) at (1.8,-.5){};
\vertex (a3) at (1.8,.5){};
\vertex (a4) at (-1.5,-1){};
\vertex (mid1) at (0,-.5);
\vertex (mid2) at (0,.5);
\vertex (mid3) at (.6,0);
\vertex (mid4) at (1.2,0);
\vertex (mid5) at (-1,.5);
\vertex (mid6) at (-1,-.5);
\vertex (mid7) at (-.5,0) {};
\diagram{
(a4) --[ultra thick,](mid6),
(a3) --[ultra thick,](mid4),
(a2) --[ultra thick,](mid4),
(a1) --[ultra thick,](mid5),
(mid1) --[ultra thick,](mid6),
(mid3) --[ultra thick,](mid2),
(mid1) --[ultra thick,](mid2),
(mid1) --[ultra thick,](mid3),
(mid5) --[ultra thick,](mid6),
(mid5) --[ultra thick,](mid2),
(mid3) --[ultra thick,](mid4),
};
\end{feynman}
\end{tikzpicture}
}
}

\newcommand{\MCutJ}{ {
\begin{tikzpicture}
\begin{feynman}
\vertex (a1) at (-1,-.5){};
\vertex (a4) at (-1,.5){};
\vertex (a2) at (2.25,-.5){};
\vertex (a3) at (2.25,.5){};
\vertex (mid1) at (-.5,0);
\vertex (mid2) at (0,0);
\vertex (mid3) at (1.25,.0) ;
\vertex (mid5) at (.25,.75);
\vertex (mid6) at (1,.75) ;
\vertex (mid4) at (1.75,.0) ;

\diagram{
(a4) --[ultra thick,](mid1),
(a3) --[ultra thick,](mid4),
(a2) --[ultra thick,](mid4),
(a1) --[ultra thick,](mid1),
(mid1) --[ultra thick,](mid2),
(mid3) --[ultra thick](mid2),
(mid3) --[ultra thick](mid6),
(mid5) --[ultra thick](mid2),
(mid5) --[ultra thick,half left](mid6),
(mid5) --[ultra thick,half right](mid6),
(mid3) --[ultra thick,](mid4),
};
\end{feynman}
\end{tikzpicture}
}
}

\newcommand{\MCutK}{ {
\begin{tikzpicture}
\begin{feynman}
\vertex (a2) at (-2.2,0){};
\vertex (a3) at (-2.2,1){};
\vertex (a1) at (-1.6,.5);

\vertex (a4) at (-1.5,-1){};
\vertex (mid1) at (0,-.5);
\vertex (mid2) at (0,.5);
\vertex (mid3) at (.6,0);
\vertex (mid4) at (1.2,0);
\vertex (mid5) at (-1,.5);
\vertex (mid6) at (-1,-.5);
\vertex (mid7) at (-.5,0) {};
\diagram{
(a4) --[ultra thick,](mid6),
(a1) --[ultra thick,](mid5),
(a1) --[ultra thick,](a2),
(a1) --[ultra thick,](a3),
(mid1) --[ultra thick,](mid6),
(mid3) --[ultra thick,](mid2),
(mid1) --[ultra thick,](mid2),
(mid1) --[ultra thick,](mid3),
(mid5) --[ultra thick,](mid6),
(mid5) --[ultra thick,](mid2),
(mid3) --[ultra thick,](mid4),
};
\end{feynman}
\end{tikzpicture}
}
}

\newcommand{\MCutL}{ {
\begin{tikzpicture}
\begin{feynman}
\vertex (a1) at (-1.5,.85){};
\vertex (a2) at (1.5,-.85){};
\vertex (a3) at (1.5,.85){};
\vertex (a4) at (-1.5,-.85){};
\vertex (mid1) at (-.25,0);
\vertex (mid2) at (.25,0);
\vertex (mid3) at (1,.4);
\vertex (mid4) at (1,-.4) ;
\vertex (mid5) at (-1,.4);
\vertex (mid6) at (-1,-.4) ;
\diagram{
(a4) --[ultra thick,](mid6),
(a3) --[ultra thick,](mid3),
(a2) --[ultra thick,](mid4),
(a1) --[ultra thick,](mid5),
(mid1) --[ultra thick,](mid2),
(mid3) --[ultra thick,](mid2),
(mid4) --[ultra thick,](mid2),
(mid4) --[ultra thick,](mid3),
(mid1) --[ultra thick,](mid6),
(mid5) --[ultra thick,](mid1),
(mid5) --[ultra thick,](mid6),
};
\end{feynman}
\end{tikzpicture}
}
}

\newcommand{\crossBoxCut}{ {
\begin{tikzpicture}
\begin{feynman}
\vertex (a1) at (-1.5,1){};
\vertex (a2) at (1.5,-1){};
\vertex (a3) at (1.5,1){};
\vertex (a4) at (-1.5,-1){};
\vertex (mid1) at (0,-.5);
\vertex (mid2) at (0,.5);
\vertex (mid3) at (1,.5);
\vertex (mid4) at (1,-.5);
\vertex (mid5) at (-1,.5);
\vertex (mid6) at (-1,-.5);
\vertex (mid7) at (-.5,0) {};
\diagram{
(a4) --[ultra thick,](mid6),
(a3) --[ultra thick,](mid3),
(a2) --[ultra thick,](mid4),
(a1) --[ultra thick,](mid5),
(mid1) --[ultra thick,](mid6),
(mid3) --[ultra thick,](mid2),
(mid1) --[ultra thick,](mid4),
(mid5) --[ultra thick,](mid7),
(mid7) --[ultra thick,](mid1),
(mid2) --[ultra thick,](mid6),
(mid5) --[ultra thick,](mid2),
(mid3) --[ultra thick,](mid4),
};
\end{feynman}
\end{tikzpicture}
}
}

\newcommand{\KTBTContrib}[4]{ {
    \begin{tikzpicture}
      \pgfmathsetmacro{\cutlength}{0.8}
\begin{feynman}
\vertex (a1) at (-1.5,1){#4};
\vertex (a2) at (1.5,-1){#2};
\vertex (a3) at (1.5,1){#3};
\vertex (a4) at (-1.5,-1){#1};
\vertex (mid1) at (-0.25,0);
\vertex (mid2) at (0.25,0);

\vertex (mid3) at (1,.5) ;
\vertex (mid4) at (1,-.5) ;
\vertex (mid5) at (-1,.5);
\vertex (mid6) at (-1,-.5) ;
\vertex (m1) at ($(mid5)!0.5!(mid6)$);
\vertex (m2) at ($(mid1)!0.5!(mid6)$);
\vertex (m3) at ($(mid5)!0.5!(mid2)$);
\vertex (clo) at ($(m1)!0.35!(mid1)$) {};
\vertex (c1) at ($(clo)+(60:\cutlength)$);
\vertex (c2) at ($(clo)+(180:\cutlength)$);
\vertex (c3) at ($(clo)+(300:\cutlength)$);
\vertex (m4) at ($(mid3)!0.5!(mid4)$);
\vertex (m5) at ($(mid3)!0.5!(mid2)$);
\vertex (m6) at ($(mid4)!0.5!(mid1)$);
\vertex (cro) at ($(m4)!0.4!(mid2)$) {};
\vertex (c4) at ($(cro)+(120:\cutlength)$);
\vertex (c5) at ($(cro)+(0:\cutlength)$);
\vertex (c6) at ($(cro)+(-120:\cutlength)$);

\diagram{
(a4) --[ultra thick,](mid6),
(a3) --[ultra thick,](mid3),
(a2) --[ultra thick,](mid4),
(a1) --[ultra thick,](mid5),
(mid1) --[ultra thick,ucp-color](mid2),
(mid3) --[ultra thick,](mid2),
(mid1) --[ultra thick,](mid6),
(mid2) --[ultra thick,](mid4),
(mid5) --[ultra thick,](mid1),
(mid5) --[ultra thick,](mid6),
(mid3) --[ultra thick,](mid4),
(clo)--[ultra thick,dashed,cut-color](c1),
(clo)--[ultra thick,dashed,cut-color](c2),
(clo)--[ultra thick,dashed,cut-color](c3),
(cro)--[ultra thick,dashed,cut-color](c4),
(cro)--[ultra thick,dashed,cut-color](c5),
(cro)--[ultra thick,dashed,cut-color](c6)
};
\end{feynman}
\end{tikzpicture}
}
}

\newcommand{\IUCContrib}[7]{ {
    \begin{tikzpicture}
      \pgfmathsetmacro{\cutlength}{0.8}
\begin{feynman}
\vertex (a1) at (-1.5,1){\scriptsize{#4}};
\vertex (a2) at (1.5,-1){\scriptsize{#2}};
\vertex (a3) at (1.5,1){\scriptsize{#3}};
\vertex (a4) at (-1.5,-1){\scriptsize{#1}};
\vertex (mid1) at (0,-.5);
\vertex (mid2) at (0,.5);
\vertex (mid3) at (1,.5) ;
\vertex (mid4) at (1,-.5) ;
\vertex (mid5) at (-1,.5);
\vertex (mid6) at (-1,-.5) ;
\vertex (m1) at ($(mid5)!0.5!(mid6)$);
\vertex (m2) at ($(mid1)!0.5!(mid6)$);
\vertex (m3) at ($(mid5)!0.5!(mid2)$);
\vertex (clo) at ($(m2)!0.5!(m3)$) {};
\vertex (c1) at ($(clo)+(90:\cutlength)$);
\vertex (c2) at ($(clo)+(180:\cutlength)$);
\vertex (c3) at ($(clo)+(270:\cutlength)$);
\vertex (m4) at ($(mid3)!0.5!(mid4)$);
\vertex (m5) at ($(mid3)!0.5!(mid2)$);
\vertex (m6) at ($(mid4)!0.5!(mid1)$);
\vertex (cro) at ($(m5)!0.5!(m6)$) {};
\vertex (c4) at ($(cro)+(90:\cutlength)$);
\vertex (c5) at ($(cro)+(0:\cutlength)$);
\vertex (c6) at ($(cro)+(270:\cutlength)$);

\diagram{
(a4) --[ultra thick,](mid6),
(a3) --[ultra thick,](mid3),
(a2) --[ultra thick,](mid4),
(a1) --[ultra thick,](mid5),
(mid1) --[ultra thick,ucp-color,edge label=\scriptsize{\(#6\)}](mid2),
(mid3) --[ultra thick,](mid2),
(mid1) --[ultra thick,](mid6),
(mid1) --[ultra thick,](mid4),
(mid5) --[ultra thick,](mid2),
(mid5) --[ultra thick,ucp-color,edge label'=\scriptsize{\(#5\)}](mid6),
(mid3) --[ultra thick,ucp-color,edge label'=\scriptsize{\(#7\)}](mid4),
(clo)--[ultra thick,dashed,cut-color](c1),
(clo)--[ultra thick,dashed,cut-color](c3),
(cro)--[ultra thick,dashed,cut-color](c4),
(cro)--[ultra thick,dashed,cut-color](c6)
};
\end{feynman}
\end{tikzpicture}
}
}

\newcommand{\halfLadder}{ {
\begin{tikzpicture}[baseline=8]
\begin{feynman}
\vertex (mid1) at (0,0);
\vertex (mid2) at (.7,0);
\vertex (mid3) at (1,0);
\vertex (mid3a) at (1.5,.4){$\cdots$};
\vertex (mid4) at (1.8,0);
\vertex (mid4a) at (2.1,0);
\vertex (a6) at (3,0){$a_n$};
\vertex (a1) at (-0.9,0){$a_1$};
\vertex (a2) at (0,.9){$a_2$};
\vertex (a3) at (.7,.9){$a_3$};
\vertex (a4) at (2.1,.9){$a_{n-1}$};
\diagram{
(mid1) -- [ultra thick](a1),
(mid1) -- [ultra thick](a2),
(mid2) -- [ultra thick](a3),
(mid4a) -- [ultra thick](a4),
(mid4a) -- [ultra thick](mid4),
(mid4a) -- [ultra thick](a6),
(mid1) --[ultra thick](mid2),
(mid2) --[ultra thick](mid3),
(mid3) --[scalar,ultra thick](mid4),
};
\end{feynman}
\end{tikzpicture}
}
}

\newcommand{\halfLadderKinetic}{ {
\begin{tikzpicture}[baseline=8]
\begin{feynman}
\vertex (mid1) at (0,0);
\vertex (mid2) at (.7,0);
\vertex (mid3) at (1,0);
\vertex (mid3a) at (1.5,.4){$\cdots$};
\vertex (mid4) at (1.8,0);
\vertex (mid4a) at (2.1,0);
\vertex (a6) at (3,0){$\mu_n$};
\vertex (a1) at (-0.9,0){$\mu_1$};
\vertex (a2) at (0,.9){$\mu_2$};
\vertex (a3) at (.7,.9){$\mu_3$};
\vertex (a4) at (2.1,.9){$\mu_{n-1}$};
\diagram{
(mid1) -- [ultra thick](a1),
(mid1) -- [ultra thick](a2),
(mid2) -- [ultra thick](a3),
(mid4a) -- [ultra thick](a4),
(mid4a) -- [ultra thick](mid4),
(mid4a) -- [ultra thick](a6),
(mid1) --[ultra thick](mid2),
(mid2) --[ultra thick](mid3),
(mid3) --[scalar,ultra thick](mid4),
};
\end{feynman}
\end{tikzpicture}
}
}

\newcommand{\JacobiOneDoubleBox}{ {
\begin{tikzpicture}
\begin{feynman}
\vertex (a1) at (-1.5,1){4};
\vertex (a2) at (1.5,-1){2};
\vertex (a3) at (1.5,1){3};
\vertex (a4) at (-1.5,-1){1};
\vertex (mid1) at (0,-.5);
\vertex (mid2) at (0,.5);
\vertex (mid3) at (1,.5) ;
\vertex (mid4) at (1,-.5) ;
\vertex (mid5) at (-1,.5);
\vertex (mid6) at (-1,-.5) ;
\diagram{
(a4) --[ultra thick,](mid6),
(a3) --[ultra thick,](mid3),
(a2) --[ultra thick,](mid4),
(a1) --[ultra thick,](mid5),
(mid1) --[ultra thick,](mid2),
(mid3) --[ultra thick,](mid2),
(mid1) --[ultra thick,](mid6),
(mid1) --[ultra thick,](mid4),
(mid5) --[ultra thick,jacobi-color](mid2),
(mid5) --[ultra thick,](mid6),
(mid3) --[ultra thick,](mid4),
};
\end{feynman}
\end{tikzpicture}
}
}

\newcommand{\JacobiOneCrossedBox}{ {
\begin{tikzpicture}
\begin{feynman}
\vertex (a1) at (-1.5,1){4};
\vertex (a2) at (1.5,-1){2};
\vertex (a3) at (1.5,1){3};
\vertex (a4) at (-1.5,-1){1};
\vertex (mid1) at (0,-.5);
\vertex (mid2) at (0,.5);
\vertex (mid3) at (1,.5);
\vertex (mid4) at (1,-.5);
\vertex (mid5) at (-1,.5);
\vertex (mid6) at (-1,-.5);
\vertex (mid7) at (-.5,0) {};
\diagram{
(a4) --[ultra thick,](mid6),
(a3) --[ultra thick,](mid3),
(a2) --[ultra thick,](mid4),
(a1) --[ultra thick,](mid5),
(mid1) --[ultra thick,](mid6),
(mid3) --[ultra thick,](mid2),
(mid1) --[ultra thick,](mid4),
(mid5) --[ultra thick,](mid7),
(mid7) --[ultra thick,](mid1),
(mid2) --[ultra thick,](mid6),
(mid5) --[ultra thick,jacobi-color](mid2),
(mid3) --[ultra thick,](mid4),
};
\end{feynman}
\end{tikzpicture}
}
}

\newcommand{\JacobiOnePentaTriangle}{ {
\begin{tikzpicture}
\begin{feynman}
\vertex (a1) at (-1.5,1){4};
\vertex (a2) at (.5,-1){2};
\vertex (a3) at (.5,1){3};
\vertex (a4) at (-1.5,-1){1};
\vertex (mid1) at (-.5,-.5);
\vertex (mid2) at (-1,0);
\vertex (mid3) at (0,.5) ;
\vertex (mid4) at (0,-.5) ;
\vertex (mid5) at (-1,.5);
\vertex (mid6) at (-1,-.5) ;
\diagram{
(a4) --[ultra thick,](mid6),
(a3) --[ultra thick,](mid3),
(a2) --[ultra thick,](mid4),
(a1) --[ultra thick,](mid5),
(mid1) --[ultra thick,](mid2),
(mid6) --[ultra thick,](mid2),
(mid1) --[ultra thick,](mid6),
(mid1) --[ultra thick,](mid4),
(mid5) --[ultra thick,](mid3),
(mid5) --[ultra thick,jacobi-color](mid2),
(mid3) --[ultra thick,](mid4),
};
\end{feynman}
\end{tikzpicture}
}
}

\newcommand{\JacobiTwoCrossedBoxOne}{ {
\begin{tikzpicture}
\begin{feynman}
\vertex (a1) at (-1.5,1){2};
\vertex (a2) at (1.5,-1){4};
\vertex (a3) at (1.5,1){3};
\vertex (a4) at (-1.5,-1){1};
\vertex (mid1) at (0,-.5);
\vertex (mid2) at (0,.5);
\vertex (mid3) at (1,.5);
\vertex (mid4) at (1,-.5);
\vertex (mid5) at (-1,.5);
\vertex (mid6) at (-1,-.5);
\vertex (mid7) at (-.5,0) {};
\diagram{
(a4) --[ultra thick,](mid6),
(a3) --[ultra thick,](mid3),
(a2) --[ultra thick,](mid4),
(a1) --[ultra thick,](mid5),
(mid1) --[ultra thick,](mid6),
(mid3) --[ultra thick,](mid2),
(mid1) --[ultra thick,jacobi-color](mid4),
(mid5) --[ultra thick,](mid7),
(mid7) --[ultra thick,](mid1),
(mid2) --[ultra thick,](mid6),
(mid5) --[ultra thick,](mid2),
(mid3) --[ultra thick,](mid4),
};
\end{feynman}
\end{tikzpicture}
}
}

\newcommand{\JacobiTwoCrossedBoxTwo}{ {
\begin{tikzpicture}
\begin{feynman}
\vertex (a1) at (-1.5,1){3};
\vertex (a2) at (1.5,-1){4};
\vertex (a3) at (1.5,1){1};
\vertex (a4) at (-1.5,-1){2};
\vertex (mid1) at (0,-.5);
\vertex (mid2) at (0,.5);
\vertex (mid3) at (1,.5);
\vertex (mid4) at (1,-.5);
\vertex (mid5) at (-1,.5);
\vertex (mid6) at (-1,-.5);
\vertex (mid7) at (-.5,0) {};
\diagram{
(a4) --[ultra thick,](mid6),
(a3) --[ultra thick,](mid3),
(a2) --[ultra thick,](mid4),
(a1) --[ultra thick,](mid5),
(mid1) --[ultra thick,](mid6),
(mid3) --[ultra thick,](mid2),
(mid1) --[ultra thick,jacobi-color](mid4),
(mid5) --[ultra thick,](mid7),
(mid7) --[ultra thick,](mid1),
(mid2) --[ultra thick,](mid6),
(mid5) --[ultra thick,](mid2),
(mid3) --[ultra thick,](mid4),
};
\end{feynman}
\end{tikzpicture}
}
}

\newcommand{\JacobiTwoCrossedBoxThree}{ {
\begin{tikzpicture}
\begin{feynman}
\vertex (a1) at (-1.5,1){1};
\vertex (a2) at (1.5,-1){4};
\vertex (a3) at (1.5,1){2};
\vertex (a4) at (-1.5,-1){3};
\vertex (mid1) at (0,-.5);
\vertex (mid2) at (0,.5);
\vertex (mid3) at (1,.5);
\vertex (mid4) at (1,-.5);
\vertex (mid5) at (-1,.5);
\vertex (mid6) at (-1,-.5);
\vertex (mid7) at (-.5,0) {};
\diagram{
(a4) --[ultra thick,](mid6),
(a3) --[ultra thick,](mid3),
(a2) --[ultra thick,](mid4),
(a1) --[ultra thick,](mid5),
(mid1) --[ultra thick,](mid6),
(mid3) --[ultra thick,](mid2),
(mid1) --[ultra thick,jacobi-color](mid4),
(mid5) --[ultra thick,](mid7),
(mid7) --[ultra thick,](mid1),
(mid2) --[ultra thick,](mid6),
(mid5) --[ultra thick,](mid2),
(mid3) --[ultra thick,](mid4),
};
\end{feynman}
\end{tikzpicture}
}
}

\newcommand{\CrossedBoxGraphSym}[4]{ {
\begin{tikzpicture}
\begin{feynman}
\vertex (a1) at (-1.5,1){#1};
\vertex (a2) at (1.5,-1){#3};
\vertex (a3) at (1.5,1){#4};
\vertex (a4) at (-1.5,-1){#2};
\vertex (mid1) at (0,-.5);
\vertex (mid2) at (0,.5);
\vertex (mid3) at (1,.5);
\vertex (mid4) at (1,-.5);
\vertex (mid5) at (-1,.5);
\vertex (mid6) at (-1,-.5);
\vertex (mid7) at (-.5,0) {};
\diagram{
(a4) --[ultra thick,](mid6),
(a3) --[ultra thick,](mid3),
(a2) --[ultra thick,](mid4),
(a1) --[ultra thick,](mid5),
(mid1) --[ultra thick,](mid6),
(mid3) --[ultra thick,](mid2),
(mid1) --[ultra thick,](mid4),
(mid5) --[ultra thick,](mid7),
(mid7) --[ultra thick,](mid1),
(mid2) --[ultra thick,](mid6),
(mid5) --[ultra thick,](mid2),
(mid3) --[ultra thick,](mid4),
};
\end{feynman}
\end{tikzpicture}
}
}

\newcommand{\ZMBasisDoubleBox}{ {
\begin{tikzpicture}
\begin{feynman}
\vertex (a1) at (-1.5,1){1};
\vertex (a2) at (1.5,-1){3};
\vertex (a3) at (1.5,1){4};
\vertex (a4) at (-1.5,-1){2};
\vertex (mid1) at (0,-.5);
\vertex (mid2) at (0,.5);
\vertex (mid3) at (1,.5);
\vertex (mid4) at (1,-.5) ;
\vertex (mid5) at (-1,.5);
\vertex (mid6) at (-1,-.5) ;
\vertex (l1) at (.5,.85){$\ell_1$};
\vertex (l2) at (-.5,.85){$\ell_2$};
\diagram{
(mid6) --[ultra thick,nhpBlue](a4),
(mid3) --[ultra thick,nhpBlue](a3),
(mid4) --[ultra thick,nhpBlue](a2),
(mid5) --[ultra thick,nhpBlue](a1),
(mid1) --[ultra thick,nhpBlue](mid2),
(mid3) --[ultra thick, fermion2,nhpBlue](mid2), 
(mid1) --[ultra thick,nhpBlue](mid6),
(mid1) --[ultra thick,nhpBlue](mid4),
(mid5) --[ultra thick, fermion2,nhpBlue](mid2), 
(mid5) --[ultra thick,nhpBlue](mid6),
(mid3) --[ultra thick,nhpBlue](mid4),
};
\end{feynman}
\end{tikzpicture}
}
}

\newcommand{\ZMBasisPentaTriangle}{ {
\begin{tikzpicture}
\begin{feynman}
\vertex (a1) at (-1.5,1){4};
\vertex (a2) at (.5,-1){2};
\vertex (a3) at (.5,1){3};
\vertex (a4) at (-1.5,-1){1};
\vertex (mid1) at (-.5,-.5);
\vertex (mid2) at (-1,0);
\vertex (mid3) at (0,.5) ;
\vertex (mid4) at (0,-.5) ;
\vertex (mid5) at (-1,.5);
\vertex (mid6) at (-1,-.5) ;
\vertex (l1) at (-1.35,.3){$\ell_1$};
\vertex (l2) at (-1.35,-.3){$\ell_2$};
\diagram{
(a4) --[ultra thick,nhpBlue](mid6),
(a3) --[ultra thick,nhpBlue](mid3),
(a2) --[ultra thick,nhpBlue](mid4),
(a1) --[ultra thick,nhpBlue](mid5),
(mid1) --[ultra thick,nhpBlue](mid2),
(mid6) --[ultra thick, fermion2,nhpBlue](mid2), 
(mid1) --[ultra thick,nhpBlue](mid6),
(mid1) --[ultra thick,nhpBlue](mid4),
(mid5) --[ultra thick,nhpBlue](mid3),
(mid5) --[ultra thick, fermion2,nhpBlue](mid2), 
(mid3) --[ultra thick,nhpBlue](mid4),
};
\end{feynman}
\end{tikzpicture}
}
}

\newcommand{\NLSMBasisDoubleBox}{ {
\begin{tikzpicture}
\begin{feynman}
\vertex (a1) at (-1.5,1){1};
\vertex (a2) at (1.5,-1){3};
\vertex (a3) at (1.5,1){4};
\vertex (a4) at (-1.5,-1){2};
\vertex (mid1) at (0,-.5);
\vertex (mid2) at (0,.5);
\vertex (mid3) at (1,.5);
\vertex (mid4) at (1,-.5) ;
\vertex (mid5) at (-1,.5);
\vertex (mid6) at (-1,-.5) ;
\vertex (l1) at (.5,.85){$\ell_1$};
\vertex (l2) at (-.5,.85){$\ell_2$};
\diagram{
(mid6) --[ultra thick, fermion2](a4),
(mid3) --[ultra thick, fermion2](a3),
(mid4) --[ultra thick, fermion2](a2),
(mid5) --[ultra thick, fermion2](a1),
(mid1) --[ultra thick,](mid2),
(mid3) --[ultra thick, fermion2](mid2), 
(mid1) --[ultra thick,](mid6),
(mid1) --[ultra thick,](mid4),
(mid5) --[ultra thick, fermion2](mid2), 
(mid5) --[ultra thick,](mid6),
(mid3) --[ultra thick,](mid4),
};
\end{feynman}
\end{tikzpicture}
}
}

\newcommand{\NLSMBasisPentaTriangle}{ {
\begin{tikzpicture}
\begin{feynman}
\vertex (a1) at (-1.5,1){4};
\vertex (a2) at (.5,-1){2};
\vertex (a3) at (.5,1){3};
\vertex (a4) at (-1.5,-1){1};
\vertex (mid1) at (-.5,-.5);
\vertex (mid2) at (-1,0);
\vertex (mid3) at (0,.5) ;
\vertex (mid4) at (0,-.5) ;
\vertex (mid5) at (-1,.5);
\vertex (mid6) at (-1,-.5) ;
\vertex (l1) at (-1.35,.3){$\ell_1$};
\vertex (l2) at (-1.35,-.3){$\ell_2$};
\diagram{
(a4) --[ultra thick,](mid6),
(a3) --[ultra thick,](mid3),
(a2) --[ultra thick,](mid4),
(a1) --[ultra thick,](mid5),
(mid1) --[ultra thick,](mid2),
(mid6) --[ultra thick, fermion2](mid2), 
(mid1) --[ultra thick,](mid6),
(mid1) --[ultra thick,](mid4),
(mid5) --[ultra thick,](mid3),
(mid5) --[ultra thick, fermion2](mid2), 
(mid3) --[ultra thick,](mid4),
};
\end{feynman}
\end{tikzpicture}
}
}

\newcommand{\ExCutLabeled}[4]{ {
\begin{tikzpicture}
\begin{feynman}
\vertex (a1) at (-.6, -0.6) {\scriptsize{#1}};
\vertex (a2) at (-.6, 0.6) {\scriptsize{#2}};
\blob{mid1}{0,0};
\blob{mid3}{.8,0};
\blob{mid5}{1.6,0};
\vertex (a3) at (2.2, .6) {\scriptsize{#3}};
\vertex (a4) at (2.2, -.6) {\scriptsize{#4}};
\diagram{
(mid1) --[ ultra thick,](a1),
(mid1) --[ ultra thick,](a2),
(mid1) --[ ultra thick,out=60,in=120,min distance=0.4cm,edge label=\scriptsize{$\ell_1$}](mid3),
(mid1) --[ ultra thick,out=-60,in=-120,min distance=0.4cm,edge label'=\scriptsize{$\ell_2$}](mid3),
(mid3) --[ ultra thick,out=60,in=120,min distance=0.4cm,edge label=\scriptsize{$\ell_3$}](mid5),
(mid3) --[ ultra thick,out=-60,in=-120,min distance=0.4cm,edge label'=\scriptsize{$\ell_4$}](mid5),
(mid5) --[ ultra thick](a4),
(mid5) --[ ultra thick,](a3)
};
\end{feynman}
\end{tikzpicture}
}
}

\newcommand{\PhysicalCutOne}[4]{ {
\begin{tikzpicture}[baseline=6]
\begin{feynman}
\vertex (a1) at (-1, 1) {#1};
\vertex (a2) at (1, 1) {#2};
\vertex [dot,scale=2.4](mid1) at (0.5,0.5){};
\vertex [dot,scale=1.8,hgrey0](mid2) at (0.5,0.5){};
\vertex [dot,scale=2.4](mid3) at (-0.5,0.5){};
\vertex [dot,scale=1.8,hgrey0](mid4) at (-0.5,0.5){};
\vertex [dot,scale=2.4](mid5) at (0,0){};
\vertex [dot,scale=1.8,hgrey0](mid6) at (0,0){};
\vertex (a3) at (-.5, -.5) {#3};
\vertex (a4) at (.5, -.5) {#4};
\diagram{
(mid3) --[ ultra thick](a1),
(mid1) --[ ultra thick](a2),
(mid1) --[ ultra thick,out=120,in=60,min distance=0.1cm](mid3),
(mid1) --[ ultra thick](mid3),

(mid1) --[ ultra thick](mid5),
(mid3) --[ ultra thick](mid5),

(mid5) --[ ultra thick](a4),
(mid5) --[ ultra thick,](a3)
};
\end{feynman}
\end{tikzpicture}
}
}

\newcommand{\PhysicalCutTwo}[4]{ {
\begin{tikzpicture}
\begin{feynman}
\vertex (a1) at (-.6, -0.6) {#1};
\vertex (a2) at (-.6, 0.6) {#2};
\vertex [dot,scale=2.4](mid1) at (0,0){};
\vertex [dot,scale=1.8,hgrey0](mid2) at (0,0){};
\vertex [dot,scale=2.4](mid3) at (.8,0){};
\vertex [dot,scale=1.8,hgrey0](mid4) at (.8,0){};
\vertex [dot,scale=2.4](mid5) at (1.6,0){};
\vertex [dot,scale=1.8,hgrey0](mid6) at (1.6,0){};
\vertex (a3) at (2.2, .6) {#3};
\vertex (a4) at (2.2, -.6) {#4};
\diagram{
(mid1) --[ ultra thick,](a1),
(mid1) --[ ultra thick,](a2),
(mid1) --[ ultra thick,out=60,in=120,min distance=0.4cm](mid3),
(mid1) --[ ultra thick,out=-60,in=-120,min distance=0.4cm](mid3),
(mid3) --[ ultra thick,out=60,in=120,min distance=0.4cm](mid5),
(mid3) --[ ultra thick,out=-60,in=-120,min distance=0.4cm](mid5),
(mid5) --[ ultra thick](a4),
(mid5) --[ ultra thick,](a3)
};
\end{feynman}
\end{tikzpicture}
}
}

\newcommand{\nGonVector}{ {
\begin{tikzpicture}[baseline=-2]
\begin{feynman}
\vertex (mid1) at (.7*.8,0);
\vertex (mid2) at (.35*.8,.606*.8);
\vertex (mid3) at (-.35*.8,.606*.8);
\vertex (mid4) at (-.7*.8,0);
\vertex (mid5) at (-.35*.8,-.606*.8);
\vertex (mid6) at (.35*.8,-.606*.8);
\vertex (a1) at (1.7,0){$n\!-\!1$};
\vertex (a2) at (.7,1.212){$n$};
\vertex (a3) at (-.7,1.212){$1$};
\vertex (a4) at (-1.4,0){$2$};
\diagram{
(mid1) -- [ultra thick,fermion2](a1),
(mid2) -- [ultra thick,fermion2](a2),
(mid3) -- [ultra thick,fermion2](a3),
(mid4) -- [ultra thick,fermion2](a4),
(mid1) --[ultra thick,fermion2](mid2),
(mid2) --[ultra thick,fermion2](mid3),
(mid3) --[ultra thick,fermion2](mid4),
(mid4) --[ultra thick,fermion2](mid5),
(mid5) --[scalar,ultra thick](mid6),
(mid6) --[ultra thick,fermion2](mid1),
}; 
\end{feynman}
\end{tikzpicture}
}
}

\newcommand{\nGonScalar}{ {
\begin{tikzpicture}[baseline=-2]
\begin{feynman}
\vertex (mid1) at (.7*.8,0);
\vertex (mid2) at (.35*.8,.606*.8);
\vertex (mid3) at (-.35*.8,.606*.8);
\vertex (mid4) at (-.7*.8,0);
\vertex (mid5) at (-.35*.8,-.606*.8);
\vertex (mid6) at (.35*.8,-.606*.8);
\vertex (a1) at (1.7,0){$n\!-\!1$};
\vertex (a2) at (.7,1.212){$n$};
\vertex (a3) at (-.7,1.212){$1$};
\vertex (a4) at (-1.4,0){$2$};
\diagram{
(mid1) -- [ultra thick,fermion2](a1),
(mid2) -- [ultra thick,fermion2](a2),
(mid3) -- [ultra thick,fermion2](a3),
(mid4) -- [ultra thick,fermion2](a4),
(mid1) --[ultra thick,nhpRed](mid2),
(mid2) --[ultra thick,nhpRed](mid3),
(mid3) --[ultra thick,nhpRed](mid4),
(mid4) --[ultra thick,nhpRed](mid5),
(mid5) --[scalar,ultra thick,nhpRed](mid6),
(mid6) --[ultra thick,nhpRed](mid1),
}; 
\end{feynman}
\end{tikzpicture}
}
}

\newcommand{\LMCut}{ {
\begin{tikzpicture}
\begin{feynman}
\vertex (a1) at (-.8, -0.6) {};
\vertex (a2) at (-.8, 0.6) {};
\vertex [dot, scale=2.4](mid1) at (0,0){};
\vertex [dot, scale=1.8,hgrey0](mid2) at (0,0){};
\vertex [dot, scale=2.4](mid3) at (1,0){};
\vertex [dot, scale=1.8,hgrey0](mid4) at (1,0){};
\vertex (a3) at ($(mid3)+(0.8, -0.6)$) {};
\vertex (a4) at ($(mid3)+(.8, 0.6)$) {};
\diagram{
(mid1) --[ultra thick,](a1),
(mid1) --[ultra thick,](a2),
(mid1) --[ultra thick,](mid3),
(mid1) --[ultra thick,out=60,in=120,min distance=0.4cm](mid3),
(mid1) --[ultra thick,out=-60,in=-120,min distance=0.4cm](mid3),
(mid3) --[ultra thick](a4),
(mid3) --[ultra thick,](a3),
};
\end{feynman}
\end{tikzpicture}
}
}

\newcommand{\simplebox}{ {
\begin{tikzpicture}
\begin{feynman}
\vertex (a1) at (-1,1){1};
\vertex (a2) at (1,-1){3};
\vertex (a3) at (1,1){4};
\vertex (a4) at (-1,-1){2};
\vertex (mid3) at (.4,.4);
\vertex (mid4) at (.4,-.4);
\vertex (mid5) at (-.4,.4);
\vertex (mid6) at (-.4,-.4);
\diagram{(mid4) --[ultra thick,fermion2](mid3),
(mid5) --[ultra thick,fermion2](mid6),
(mid3) --[ultra thick,fermion2](mid5),
(mid6) --[ultra thick,fermion2](mid4),
(mid6) --[ultra thick,fermion2](a4),
(mid3) --[ultra thick,fermion2](a3),
(mid4) --[ultra thick,fermion2](a2),
(mid5) --[ultra thick,fermion2](a1),
};
\end{feynman}
\end{tikzpicture}
}
}

\newcommand{\propFrule}[4]{ {
\begin{tikzpicture}
\begin{feynman}
\vertex (a2) at (1,0){#2};
\vertex (a3) at (-1,0){#1};
\vertex (mid1) at (0,.3){#3};
\diagram{
(a3) --[ultra thick,#4](a2),
};
\end{feynman}
\end{tikzpicture}
}
}

\newcommand{\cubicFrule}[6]{ {
\begin{tikzpicture}
\begin{feynman}
\vertex (a1) at (0,1){#2};
\vertex (a2) at (1,0){#3};
\vertex (a3) at (-1,0){#1};
\vertex (mid1) at (0,0);
\diagram{
(mid1) --[ultra thick,#4](a3),
(mid1) --[ultra thick,#5](a2),
(a1) --[ultra thick,#6](mid1),
};
\end{feynman}
\end{tikzpicture}
}
}

\newcommand{\cubic}[7]{ {
\begin{tikzpicture}
\begin{feynman}
\vertex [dot, scale=2.6](mid1) at (0,0){};
\vertex [dot, scale=2.6](mid2) at (0,0){};
\vertex [dot, scale=2,#1](mid3) at (0,0){};
\vertex (a1) at (0,1){3};
\vertex (a2) at (.85,-.55){2};
\vertex (a3) at (-.85,-.55){1};
\diagram{
(mid1) --[ultra thick,#2,#3](a3),
(mid1) --[ultra thick,#4,#5](a2),
(a1) --[ultra thick,#6,#7](mid1),
};
\end{feynman}
\end{tikzpicture}
}
}

\newcommand{\Acubic}[7]{ {
\begin{tikzpicture}
\begin{feynman}
\vertex (a1) at (0,1){3};
\vertex (a2) at (.85,-.55){2};
\vertex (a3) at (-.85,-.55){1};
\vertex [dot, scale=2.6](mid1) at (0,0){};
\vertex [dot, scale=2.6](mid2) at (0,0){};
\vertex [dot, scale=2,#1](mid3) at (0,0){};
\diagram{
(a3) --[ultra thick,#2,#3](mid1),
(a2) --[ultra thick,#4,#5](mid1),
(mid1) --[ultra thick,#6,#7](a1),
};
\end{feynman}
\end{tikzpicture}
}
}

\newcommand{\eg}{e.g.,}
\usepackage{xspace}

\def\spa#1.#2{\left\langle#1\,#2\right\rangle}
\def\spb#1.#2{\left[#1\,#2\right]}
\def\spash#1.#2{\spa{\smash{#1}}.{\smash{#2}}}
\def\spbsh#1.#2{\spb{\smash{#1}}.{\smash{#2}}}
\def\sand#1.#2.#3{%
\left\langle\smash{#1}{\vphantom1}^{-}\right|{#2}%
\left|\smash{#3}{\vphantom1}^{-}\right\rangle}
\def\sandpp#1.#2.#3{%
\left\langle\smash{#1}{\vphantom1}^{+}\right|{#2}%
\left|\smash{#3}{\vphantom1}^{+}\right\rangle}
\def\sandpm#1.#2.#3{%
\left\langle\smash{#1}{\vphantom1}^{+}\right|{#2}%
\left|\smash{#3}{\vphantom1}^{-}\right\rangle}
\def\sandmp#1.#2.#3{%
\left\langle\smash{#1}{\vphantom1}^{-}\right|{#2}%
\left|\smash{#3}{\vphantom1}^{+}\right\rangle}

\def\be{\begin{equation}}
\def\ee{\end{equation}}
\def\bea{\begin{eqnarray}}
\def\eea{\end{eqnarray}}
\def\ba{\begin{eqnarray}}
\def\ea{\end{eqnarray}}

\newcommand{\ansatze}{ans\"atze}

\newcommand{\atree}{\ensuremath{A^{\text{tree}}}}

\definecolor{NUpurple}{RGB}{078,042,132}

\author{\large Alex Edison \orcidlink{0000-0002-5430-9500},}
\author{\large James Mangan  \orcidlink{0000-0002-9713-7446},}
\author{\large and Nicolas H. Pavao \orcidlink{0000-0002-9817-8266}}

\affiliation{Department of Physics and Astronomy, Northwestern
  University, Evanston, Illinois 60208, USA}

\emailAdd{alexander.edison@northwestern.edu, 
\\
james.mangan@northwestern.edu, 
\\
pavao@u.northwestern.edu }
  
\title{\center  \fontsize{18}{20} \selectfont  Revealing the Landscape of Globally Color-Dual Multi-loop Integrands}

\abstract{We report on progress in understanding how to construct color-dual multi-loop amplitudes.
First we identify a cubic theory, semi-abelian Yang-Mills, that unifies
  many of the color-dual theories studied in the literature, and
  provides a prescriptive approach for constructing $D$-dimensional color-dual
  numerators through one-loop directly from Feynman rules. By a simple weight counting argument, this approach does not further generalize to two-loops.  As a first step in
  understanding the two-loop challenge, we use a $D$-dimensional color-dual bootstrap to 
  successfully construct globally color-dual
  \emph{local} two-loop four-point nonlinear sigma model (NLSM) numerators.  The double-copy of
these NLSM numerators with themselves, pure Yang-Mills, and
  $\mathcal{N}=4$ super-Yang-Mills correctly reproduce the known unitarity constructed integrands 
  of special Galileons, Born-Infeld theory,
  and Dirac-Born-Infeld-Volkov-Akulov theory, respectively. 
  Applying our bootstrap to two-loop four-point pure Yang-Mills, we exhaustively search the space of local numerators and find that it fails to satisfy global color-kinematics duality, completing a search previously initiated in the literature. 
  We pinpoint the failure to the
  bowtie unitarity cut, and discuss a path forward towards \emph{non-local}
  construction of color-dual integrands at generic loop order.
  
}

\makeatletter
\gdef\@fpheader{\,}

\begin{document}
\maketitle
\flushbottom
 
\setstackgap{S}{6pt}
\setstackgap{L}{7pt}

\section{Introduction}\label{intro}
Since the turn of the century, our understanding of the $S$-matrix and
its concealed structures has expanded dramatically. At the heart of this
progress is the mantra that physical observables are simpler when
studied on-shell
\cite{Parke:1986gb,TasiLance,Cheung:2017pzi}. However, while inserting on-shell states certainly offers a path towards taming the
factorial growth of Feynman diagrams, it also obscures the off-shell
simplicity at the heart of many quantum field theories. A shining
example of this hidden structure is the duality between color and
kinematics \cite{BCJ,Bern:2010ue,BCJreview}, which states that the
kinematic numerators of gauge theory amplitudes can be rearranged to
obey the same algebraic identities as the constituent color
factors. When this duality is realized globally off-shell, it
dramatically decreases the combinatorial complexity posed by integrand
construction, and offers a path towards efficient assembly of quantum
gravity integrands directly from simpler gauge theory building blocks
via the double copy \cite{BCJ,Bern:2010ue}.

Despite the tremendous success in leveraging this duality to
compute gauge and gravity observables to high orders in perturbation
theory \cite{FiveLoopN4, GeneralizedDoubleCopyFiveLoops, Bern:2018jmv}, color-kinematics remains a conjecture at loop
level. Indeed, there are many examples in the literature where
identifying color-dual representations beyond one-loop has posed a
formidable challenge
\cite{Mogull:2015adi, Johansson:2017bfl, KalinN2TwoLoop, Bern:2015ooa}. In this work, we use
the nonlinear sigma model (NLSM) and Yang-Mills, two theories proven
to permit color-dual representations at tree-level
\cite{Feng:2010my,Cachazo:2014xea}, as case studies to advance our
understanding of the kinematic algebra at the multi-loop level.

We begin with an overview of color-kinematics duality in
\cref{background}, and define the notion of ``globally'' color-dual
integrands in \cref{sec:bootstrap}.  We then construct a manifestly
color-dual theory in \cref{sec:cubic}, which generates $D$-dimensional
color-dual $n$-point numerators at both tree-level and one-loop. When
plugging in appropriate on-shell states, we find that these numerators
underpin color-dual representations of self-dual Yang-Mills, NLSM and
Chern-Simons theory through one-loop. The construction of this theory,
which we dub \textit{semi-abelian Yang-Mills}, relies on isolating the
cubic sector of pure Yang-Mills amplitudes from the four-point
contacts that are needed to fully realize non-abelian gauge
symmetry. Despite the potency of this theory through one-loop, the
construction runs into an obstruction at two-loop, which we describe
in \cref{2LoopObstruction}.

Faced with this obstruction, we study the multi-loop sector of both
NLSM and Yang-Mills in \cref{2loopBoot} using an ansatz-based
color-dual bootstrap. In contrast to much of the available literature
on color-dual integrand construction at two-loop
\cite{Bern:2013yya,Mogull:2015adi,Johansson:2017bfl}, our results are
completely agnostic to the spacetime dimension; all the polarizations
and momenta appearing in our construction will remain formally
$D$-dimensional. This makes our methods and results particularly well
suited for algorithmically extracting rational terms from the
$D$-dependence of dimensionally regulated loop momenta.
The ansatz approach to constructing numerators generally results in an explosion of terms, but in the case of scalar theories there is an added difficulty because the linear equations become dense.
We were able overcome this barrier by employing a custom
solver, \texttt{FiniteFieldSolve}, which will soon be made
public.

With this solver, we are able to compute a two-loop integrand for NLSM
that globally manifests the duality off-shell. This represents a new
benchmark for color-dual representations in non-supersymmetric gauge
theories.  However, in sharp contrast, we find that Yang-Mills does
\emph{not} permit a globally color-dual representation, even when
considering the most general polynomial ansatz of Lorentz covariant
kinematics.\footnote{Globally color-dual local numerators for two-loop
  pure YM have been studied before but only in the context of a
  limited ansatz subject to certain loop power counting constraints
  \cite{Bern:2015ooa}.}  Concretely, the failure point can be
pinpointed to a conflict between the ``bowtie'' cut and the
following Jacobi triple:
\begin{equation}
   \NMCutC
  \Rightarrow
  \JacobiOneDoubleBox +  \JacobiOneCrossedBox + \JacobiOnePentaTriangle \neq 0 \,.
\end{equation}
We comment on the implications of this finding in \cref{conclusions}
and discuss alternative approaches where one must relax field
theoretic locality in order to accommodate the construction of
globally color-dual integrands at generic loop order.

\section{Background}\label{background}
To set the stage, in this section we provide an overview of
color-kinematics duality -- both the on-shell and off-shell
constructions. We then detail a color-dual bootstrap approach for
constructing integrands in theories for which the fully off-shell kinematic
algebra has not yet been identified, like Yang-Mills and NLSM.
\subsection{Color-dual Amplitudes}\label{onShellCK}

Color-kinematics duality is a statement about the off-shell graphical
simplicity that manifests redundancy in the on-shell observables of
certain field theories. Ignoring factors of $i$ and the coupling
constant, the starting point to realize the duality is to express generic $n$-point $L$-loop amplitudes in a cubic graph representation:
\begin{equation}
  \mathcal{A}_n^L = \sum \limits_\Gamma \frac{1}{S_\Gamma} \int \frac{d^{LD}\ell}{(2\pi)^{LD}} \frac{C_\Gamma N_\Gamma}{d_\Gamma} .
  \label{eq:gen-amp}
\end{equation}
Here the spacetime dimension is $D$.  Even though the underlying
theory may have quartic (or higher multiplicity) vertices, like YM and NLSM, it is
possible to multiply and divide by propagators so that the sum runs
over the set of purely cubic graphs, $\Gamma$.  Overcounting is removed by
the internal symmetry factor $S_\Gamma$, which tracks the number of
graph symmetries with external legs held fixed.  

The color factor
$C_\Gamma$ is the string of structure constants $f^{abc}$ associated
with the graph $\Gamma$,\footnote{This paper is focused on theories
  with particles that transform in the adjoint.  However, the double
  copy can be applied to a far more diverse set of theories with
  particles that transform in the fundamental or more diverse
  representations, see for example \cite{Johansson:2017srf,
    Johansson2014zca, Johansson:2015oia, Johansson:2019dnu, Carrasco:2020ywq,Carrasco:2022jxn,
    Carrasco:2023vjg}.}  the denominator $d_\Gamma$
is the product of propagators associated with the graph $\Gamma$, and
the numerator $N_\Gamma$ is all of the remaining kinematic factors
including dot products of momenta and polarization vectors. In this 
form the color factors $C_\Gamma$ are not independent and are related to
one another via a set of Jacobi identities that look schematically
like
\begin{equation}
C_i+C_j+C_k=0,
\end{equation}
where $i$, $j$, and $k$ are three graphs.  Example Jacobi relations
are presented in detail in \cref{sec:bootstrap}. The duality between color and kinematics is the statement that that there exists a way to rearrange factors between
the various kinematic numerators $N_\Gamma$ such that the numerators
obey the same Jacobi identities as the color factors
\begin{equation}
N_i +N_j +N_k=0.
\end{equation}
If color-kinematics duality can be achieved at the off-shell level,
then the kinematic numerators should satisfy an algebra just like the
color factors.  In theories where the kinematic algebra is known, it
is often the algebra of volume preserving diffeomorphisms
\cite{Monteiro2011pc, Ben-Shahar:2021zww, Cheung:2020djz,
  Cheung:2021zvb, Cheung:2022mix}.
When the kinematic algebra is not known explicitly, the kinematic numerators are typically constructed from an ansatz.
As an example, the result of such a calculation for four-point NLSM scattering at tree level produces the numerators
\begin{align}\label{pionNums}
&N^{\text{NLSM}}_s = s(t-u)\\
&N^{\text{NLSM}}_t = t(u-s)\\
&N^{\text{NLSM}}_u = u(s-t)
\end{align}
where the coupling has been normalized away.  These numerators sum to
zero by explicit calculation.  Note that the $s$-channel numerator
contains an explicit factor of $s$ that cancels with the propagator
and the same goes for the $t$- and $u$-channels.  The net result is
that the amplitude is a local function, just as one would expect for
the pion four-point tree amplitude.

Once color-dual numerators $N_\Gamma$ have been found, they can be
used to replace the color factors ($C_\Gamma \to N_\Gamma$) in a
\emph{different} amplitude to produce the corresponding amplitude in
a new theory,
\be\label{DCconstruction}
\tilde{\mathcal{A}}_n^L = \sum \limits_\Gamma \frac{1}{S_\Gamma} \int
\frac{d^{LD}\ell}{(2\pi)^{LD}} \frac{C_\Gamma
  \tilde{N}_\Gamma}{d_\Gamma} ~~ \xrightarrow{(C_\Gamma \to N_\Gamma)}
~~ \mathcal{M}_n^L = \sum \limits_\Gamma \frac{1}{S_\Gamma} \int
\frac{d^{LD}\ell}{(2\pi)^{LD}} \frac{N_\Gamma
  \tilde{N}_\Gamma}{d_\Gamma}.
\end{equation}
This is known as the double-copy construction
\cite{BCJ,Bern:2010ue,BCJreview}.  Importantly, the $\tilde{N}_\Gamma$
do not need to respect color-kinematics duality.  The prototypical
example of the double-copy is that YM double copied with YM results in
gravity where the gauge invariance of each separate gluon produces the
diffeomorphism invariance of the graviton.  The double copy can be
proven at tree level using the Kawai-Lewellen-Tye (KLT) relations but
there is significant evidence that the double copy persists to all
loop orders \cite{FiveLoopN4, GeneralizedDoubleCopyFiveLoops,
  Bern:2018jmv, KLT, KiermaierTalk}.  In addition to pointing at
undiscovered structure hidden in the Lagrangian for gravity, the
double copy is immensely useful at a utilitarian level since it is
much more efficient at producing $\mathcal{M}_n^L$ than traditional
methods like Feynman rules.

Many of the one-loop results presented in \cref{sec:cubic} are informed
by the tree-level double-copy so we will review the tree case here.
The color structure of tree-level amplitudes is particularly simple.
A fully color dressed tree amplitude $\mathcal{A}_n$ can be decomposed
in terms of the trace basis
\begin{equation}
  \mathcal{A}_n = \sum \limits_{\sigma\in S_{n-1}} \text{Tr}(T^{a_{\sigma_1}} T^{a_{\sigma_2}}...T^{a_{\sigma_{n-1}}} T^{a_n}) A_n[\sigma_1, \sigma_2,...\sigma_{n-1}, n],
  \label{eq:trace-basis}
\end{equation}
where $T^a$ are generators of the gauge group and, because of the
cyclicity of the trace, one of the legs is held fixed so that the sum
runs over the $(n-1)!$ permutations of the remaining external legs.
The coefficients of the color factors are the color-ordered partial
amplitudes $A_n[...]$.  Any tree-level color factor can be converted
into a linear combination of Del Duca-Dixon-Maltoni (DDM) half-ladder
color factors
\begin{equation}
\text{DDM}[a_1, a_2,... a_n] \equiv f^{a_1 a_2 b_1} f^{b_1 a_3 b_2} f^{b_2 a_4 b_3}...f^{b_{n-3} a_{n-1} a_n}
\end{equation}
associated with the graph
\begin{equation}
\halfLadder
\end{equation}
by repeated application of the Jacobi identity \cite{DixonMaltoni}.
The fully color-dressed tree amplitude can then be re-expressed as
\begin{equation}
  \mathcal{A}_n = \sum \limits_{\sigma\in S_{n-2}} \text{DDM}[a_1, a_{\sigma_2}, a_{\sigma_3}, a_{\sigma_{n-1}}, a_n]  A_n[1, \sigma_2,...\sigma_{n-1}, n]
  \label{eq:ddm-basis}
\end{equation}
with legs $1$ and $n$ fixed.  Since the sum only contains $(n-2)!$
terms, there must be additional relations -- the
Kleiss-Kuijf (KK) relations \cite{Kleiss:1988ne} -- amongst the
partial amplitudes appearing in the right-hand side of
\cref{eq:trace-basis}.  The KK relations are generic to any theory
with purely adjoint particles.  Color-kinematics duality further
implies that the KK basis is overcomplete and can be reduced to a
basis of $(n-3)!$ amplitudes via the fundamental Bern-Carrasco-Johansson (BCJ) identities
\cite{BCJ,Feng:2010my}
\begin{equation}
\sum \limits_{i=2}^{n-1} k_1 \cdot (k_2+...+k_i) A_n[2,...,i,1,i+1,...,n] =0.
\end{equation}
In terms of constructing explicit color-dual solutions at tree level,
it is enough to specify the numerator of the half-ladder because any
diagram can be reduced to a half-ladder through successive
applications of the Jacobi identity.  At one-loop, every diagram can
be related to one topology as well, in this case the $n$-gon, which
can be understood as a forward limit of the half-ladder.  Two-loop
calculations play an important role in our understanding of
color-kinematics duality because this is the first order where
multiple basis graphs are required, at least for a generic theory. We
refer the reader to Ref.~\cite{BCJreview} for more background on the
topic.

\subsection{Color-dual Lagrangians}\label{offShellCK}
While the BCJ relations are an on-shell statement about the color-dual
nature of scattering amplitudes, one might aspire to trivialize
the duality by encoding it in an off-shell Lagrangian
of the theory. The double-copy construction of
\cref{DCconstruction} would suggest a manifestly cubic Lagrangian of
the form
\begin{equation}\label{cubicCKLag}
\mathcal{L} = P_V^{ab}P_W^{\mu\nu}\mathcal{O}^a_\mu \Box \mathcal{O}^b_\nu +  V^{abc} W^{\mu \nu \rho}\mathcal{O}^a_\mu \mathcal{O}^b_\nu \mathcal{O}^c_\rho
\end{equation}
where $V^{abc}$ and $W^{\mu \nu \rho}$ are cubic Feynman rules mixing field
operators $\mathcal{O}^{a}_\mu$ indexed by quantum numbers $a$ and
$\mu$. The Feynman rules for this theory are simply:
\be\label{CKFrules}
\begin{split}
\propFrule{$a,\mu$}{$b,\nu$}{$k \rightarrow $}{} &= \frac{i}{k^2} (P_V^{-1})^{ab}(P_W^{-1})^{\mu\nu}
\\
\cubicFrule{$a,\mu$}{$b,\nu$}{$c,\rho$}{}{}{} &= i V^{abc}W^{\mu \nu \rho}
\end{split}
\ee
We have introduced $P_V^{ab}$ and $P_W^{\mu\nu}$ projection operators
to encode non-local kinetic structure that could in principle
participate in the construction, as is the case for $DF^2$ theory \cite{Johansson:2017srf}. Generally, these quadratic dressings are simply delta functions, $P_V^{ab} = \delta^{ab}$, or flat space metrics, $P_W^{\mu\nu}=\eta^{\mu\nu}$. As written, the theory will be color dual if the off-shell cubic vertex is antisymmetric and satisfies the Jacobi identity. Specifically,
\begin{align}
\text{antisymmetry}:& \quad{}_{\mu_1}\langle W^{\mu_2} \rangle_{\mu_3}+\text{cyc}(23)=0
\\
\text{Jacobi identity}:& \quad {}_{\mu_1}\langle W^{\mu_2} W^{\mu_3}\rangle_{\mu_4}+\text{cyc}(234)=0\label{jacID}
\end{align}
where, using the Feynman rules of \cref{CKFrules}, the bracketed expression is the half-ladder numerator of the following diagram,
\be
{}_{\mu_1}\langle W^{\mu_2}W^{\mu_3}\cdots W^{\mu_{n-1}} \rangle_{\mu_n} = N\left[\halfLadderKinetic\right]\,,
\ee
and likewise for the $V^{abc}$ dressing.
Off-shell descriptions of
color-dual theories are rare, and only a few examples are available in
the literature \cite{Monteiro2011pc, Cheung:2016prv, Cheung:2021zvb,
  Cheung:2020djz, Ben-Shahar:2022ixa, Ben-Shahar:2021zww,
  Ben-Shahar:2021doh}.
Typically, when one goes about constructing color-dual
theories directly from a set of cubic interactions, higher point
Jacobi and antisymmetry constraints require introducing additional operators \cite{Carrasco:2022lbm,Carrasco:2022sck} or propagating fields \cite{Ben-Shahar:2022ixa}.
Indeed, as we will show in \cref{sec:cubic}, Yang-Mills
requires introducing a cubic two-form interaction in order to manifest
the kinematic algebra just at four-point. Absent information about the
kinematic algebra, one can make progress in the construction of
loop-level color-dual amplitudes by employing a color-dual bootstrap.

\subsection{Color-dual Bootstrap}
\label{sec:bootstrap}

When an off-shell color-dual Lagrangian is not known, the loop-level
amplitude integrand must be constructed from an ansatz.  The four
conditions we impose on an integrand ansatz are summarized here and
elaborated on below.\footnote{The four ansatz conditions are similar
  to those of Ref.~\cite{Bern:2015ooa} except that there the authors
  demand that the numerators have the same loop power counting as
  Feynman gauge Feynman rules.  This is an additional assumption that
  is most emphatically not made here.}
After imposing the four constraints below, any remaining coefficients in the ansatz encode ``generalized gauge freedom'', meaning that the
choice of coefficients does not affect the physical integrand or its double copy.
\begin{enumerate}
\item \textbf{Off-shell Locality}: Diagram numerators are polynomials
  in momenta and polarizations, e.g., the diagrams only have poles given
  by the propagators of the diagrams.
\item \textbf{Color-kinematics duality}: The numerators obey the same
  Jacobi identities as the color factors.  This implicitly requires
  that the integrand is expressed in terms of purely cubic diagrams.
\item \textbf{Graph symmetries}: A diagram's numerator is only a
  function of the diagram's topology and labeling.  Furthermore, the
  diagram numerators are invariant under the automorphisms of the
  diagrams including signs that compensate for color-factor sign
    changes.
\item \textbf{On-shell Unitarity}: The cuts of the ansatz must
  reproduce the physical unitarity cuts of the theory when internal
  momenta are taken on-shell.
\end{enumerate}

In order to clarify general statements, the following subsections
include examples from a four-point two-loop integrand for some generic
color-dual theory like YM.  This is the simplest case that
demonstrates the full machinery of the color-dual bootstrap.  All tree
and one-loop processes are generated by single basis diagrams (the half
ladder and $n$-gon respectively) so they lack examples of three-term
``boundary'' Jacobi relations to be described shortly.
Choosing the four-point two-loop integrand also has the advantage that this
process appears prominently in this paper.

\paragraph{Off-shell Locality}

The kinematic numerators must be $D$-dimensional, Lorentz invariant,
local (polynomial) functions of the graph kinematics with the correct
power counting and external states. Locality is physically motivated since Feynman
rules produce polynomial numerators. Indeed, the most natural way to guarantee locality and avoid spurious poles is to require the off-shell numerators to be polynomial functions of
kinematics. However, insisting on a local numerator is
partly a matter of convenience as the ansatz for a rational function
would quickly grow out of control without a guiding principle for what
terms to include.  While these
assumptions are well motivated from a physical and complexity
standpoint, the literature has featured numerators that violate
locality, modify naive power counting through the inclusion of higher
spin modes, and break manifest Lorentz invariance \cite{Square,
  WeinzierlBCJLagrangian, Mogull:2015adi, FivePointN4BCJ,
  Johansson:2017bfl, Ben-Shahar:2022ixa, Cheung:2016prv,
  Cheung:2021zvb}.  These developments suggest that relaxing locality
is a viable alternative when the condition proves too stringent for an
ansatz.  However, simply requiring on-shell locality may suffice.
This alternative will be discussed in \cref{nonLocalScattering}.

\paragraph{Color-kinematics duality}

\begin{figure}
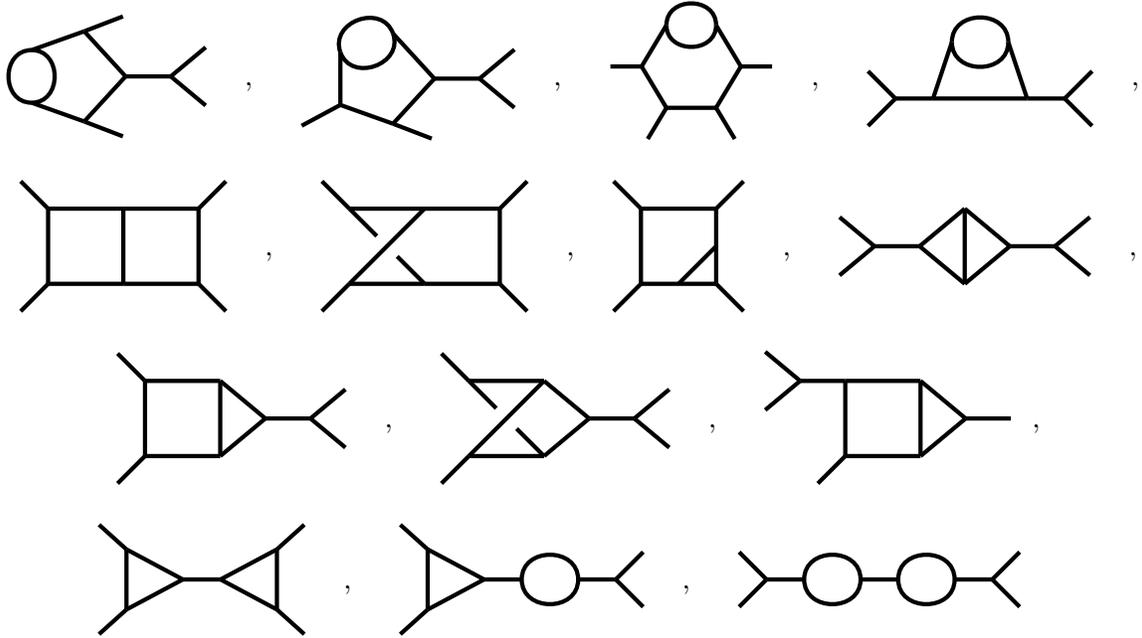

\centering
{\MCutC, \MCutD, \sumoCut, \MCutJ,  
 \\
  \doubleBoxCut,  \crossBoxCut,\pentaTriangleCut,  \MCutF, 
\\
\MCutI,  \MCutE, \MCutK, 
\\
 \MCutL, \MCutG,  \MCutH
}
\caption{The 14 non-pathological four-point two-loop cubic diagrams.} \label{fig:MaxCuts}
  \end{figure}

Color-kinematics duality is a statement about color factors, kinematic
numerators, and the accompanying Jacobi relations.  Adjoint-type color
factors are associated with cubic graphs, so satisfying
color-kinematics duality implicitly requires expressing the integrand
in terms of such graphs. As mentioned earlier, this is possible to do
even for a theory with quartic (or higher multiplicity) vertices by
multiplying and dividing by propagators.  In the case of the four-point
two-loop example promised above, there are 14 cubic graph topologies
(shown in \cref{fig:MaxCuts}) ignoring tadpoles and bubble on external
leg (BEL) graphs.

With the integrand expressed in terms of cubic graphs, it is possible
to discuss the main ingredient in color-kinematics duality -- the
Jacobi relations.  The color factors in an integrand obey a set of
Jacobi identities.  Color-kinematics duality states that there
\emph{exists} a way of writing the kinematic numerators of the
integrand so that they obey the same set of Jacobi identities.  The
Jacobi identities can be divided into two categories: \emph{defining}
relations and \emph{boundary} relations.  Defining Jacobi relations
can be used to express any graph in terms of a basis.  For example,
the double box, crossed box, and penta-triangle are related to each
other as follows
\begin{equation}
\label{eq:TwoLoopJacobiOne}
\JacobiOneDoubleBox + \JacobiOneCrossedBox + \JacobiOnePentaTriangle
=0\,,
\end{equation}
where the shared propagator is shown in red.  Using this relation the
crossed box can be written (or \emph{defined}) in terms of the other
two.  In this way, all of the 14 cubic four-point two-loop topologies can be
written in terms of a basis of the double box and penta-triangle.
(Any two of the double box, crossed box, and penta-triangle would
suffice for a basis but the planar basis is chosen for simplicity in
this work.)  The numerator of each of the two basis graphs receives
its own ansatz and the numerator of any other graph can be expressed
in terms of these two.

After using the defining Jacobi relations to relate every graph to the
basis, any remaining Jacobi relations will be referred to as boundary
relations.  In other words, boundary relations (indirectly) relate
basis elements to themselves.  For example, the crossed box numerator
is related to itself via the following Jacobi relation
\begin{equation}
\label{eq:TwoLoopJacobiTwo}
\JacobiTwoCrossedBoxOne + \JacobiTwoCrossedBoxTwo +  \JacobiTwoCrossedBoxThree =0,
\end{equation}
where the shared propagator is shown in red as before.  Each of the
crossed box numerators can be written in terms of the double box and
penta-triangle so this Jacobi relation places constraints on the basis
numerator \ansatze{}.  While defining Jacobi relations simplify how
many numerators require \ansatze{}, boundary relations place
restrictions on those \ansatze{}.  In principle there could be Jacobi
relations involving tadpoles and BEL graphs, which would be subtle to
handle.  One option is to force the numerators of these graphs to
zero.  However, since these graphs will not contribute to physical
unitarity cuts, we take a different approach.  These graphs are
allowed to have non-zero numerators but any Jacobi relations involving
them are simply not imposed or solved.  See Refs.~\cite{Bern:2012uf,
  Edison:2022jln} and \cref{BELreg} for discussions of BELs in the context of
color-kinematics duality.

What we have just described is
sometimes referred to as ``global'' color-kinematics duality because
the Jacobi identities are solved by a single (global) integrand with
off-shell internal lines.  In order to perform the double copy it is
enough to simply satisfy the Jacobi identities on the generalized
unitarity cuts, meaning that each graph can be assigned its own ansatz
\cite{Bern:2015ooa}.  If a global solution cannot be found, this is
one way to enlarge the ansatz, where another option is to incorporate non-local terms in the ``numerators''.  A globally
color-dual integrand is desirable because of its simplicity and
because a color-dual Lagrangian would naively produce precisely this
object.

\paragraph{Graph symmetries}

Every numerator is required to respect all of the symmetries (or
automorphisms) of its associated graph.  Graph symmetries are an
avatar of Bose symmetry.  For a pure ansatz construction it is very
reasonable to impose graph symmetries, but color-kinematics duality at
tree-level is often achieved at the expense of manifest Bose symmetry
\cite{Cheung:2016prv, Cheung:2021zvb, Brandhuber:2021bsf,
  Edison:2020ehu}.  In an ansatz construction graph symmetries are
imposed for \emph{every} graph, not just the basis graphs.
For example, the symmetries of the crossed box are spanned by,
\begin{equation}
\scalebox{0.8}{\CrossedBoxGraphSym{1}{2}{3}{4}}\!\! \sim\!\! \scalebox{0.8}{\CrossedBoxGraphSym{2}{1}{3}{4}}
\text{ and }
\scalebox{0.8}{\CrossedBoxGraphSym{1}{2}{3}{4}} \!\!\sim\!\! \scalebox{0.8}{\CrossedBoxGraphSym{2}{1}{4}{3}}
\end{equation}
which are just functional relabelings of external legs,
$\{1234\} \to \{2134\}$ and $\{1234\} \to \{1243\}$.  In general,
internal edge labels must be tracked as well.  Furthermore, care must
be taken with the signs in a symmetry relation.  To compensate for the
antisymmetric color factors, $f^{abc}$, graph vertices are all totally
antisymmetric and hence have signs built into their orientations.
When used in conjunction with the Jacobi relations, \eg{}
\cref{eq:TwoLoopJacobiOne}, the symmetry constraints of non-basis
graphs still impose conditions on the basis ansatz.  In fact, since
the symmetry properties of a graph are directly linked with
transformations of its color dressing, the resulting constraints can
be thought of as a subclass of the \emph{boundary} Jacobi relations.

\paragraph{On-shell Unitarity}

The final constraints come from ensuring that the ansatz correctly
reproduces the generalized unitarity cuts of the desired theory, thus
guaranteeing that the numerators encode the correct physical,
theory-dependent information.  Generalized unitarity has been
reviewed in many places
\cite{BCFUnitarity, Bern:2015ooa, Bourjaily:2017wjl,
  JJHenrikReview, FivePointN4BCJ, Bern:2012uf, BCJreview,
  Edison:2022jln, Edison:2022smn} and was recently optimized for effective field theories like NLSM \cite{Carrasco:2023qgz}. Generalized unitarity equates products of on-shell tree amplitudes,
encoded via a graph $\gamma$, to sums over compatible\footnote{A
  diagram $g$ is compatible with the cut diagram $\gamma$ if $g$ is
  isomorphic to $\gamma$ or to any of the additional factorization
  channels of $\gamma$. } diagram numerators evaluated on the support
of the on-shell conditions and weighted by their uncut propagators
\begin{equation}
\label{eq:SchematicCutEquation}
\text{Cut}[\gamma]= \sum_{\substack{\text{states}\\\text{crossing } E(\gamma)}} \prod_{v \in V(\gamma)} \atree(v)
= \sum_{g \text{ compat } \gamma}
\frac{N[g] \vert_\text{cut}}{\prod \limits_{\text{uncut}} \ell^2}
 \, ,
\end{equation}
where $V(\gamma)$ are the vertices of $\gamma$ and $E(\gamma)$ its
edges.  The sum over states hides much of the complexity of cut
construction: for scalars the process is trivial while spinning states
greatly complicate matters (see
Refs.~\cite{Kosmopoulos:2020pcd,Edison:2022jln} for recent discussion
and detailed examples of evaluating state sums).  As an example of the
two types of expressions appearing in \cref{eq:SchematicCutEquation},
the iterated two-particle cut can be
represented as a product of trees via
\begin{equation}
  \label{eq:example-cut-prod}
  \ExCutLabeled{1}{2}{3}{4} = \!\!\!\!\! \sum_{\substack{\text{states of }\\ \{\ell_1, \ell_2, \ell_3, \ell_4\}}} \!\!\!\!\!\!\!
  \atree(1,2,\ell_1,\ell_2) \atree(-\ell_1,-\ell_2,\ell_3,\ell_4) \atree(-\ell_3,-\ell_4,3,4)
\end{equation}
or as a collection of diagram numerators via
\begin{align}
  \label{eq:example-cut-sum}
  \ExCutLabeled{1}{2}{3}{4} &=
  \frac{\delta(\textcolor{cut-color}{\ell_1^2})
    \delta(\textcolor{cut-color}{\ell_2^2})
    \delta(\textcolor{cut-color}{\ell_3^2})
    \delta(\textcolor{cut-color}{\ell_4^2})}{(\textcolor{ucp-color}{\ell_5})^2
    (\textcolor{ucp-color}{\ell_6})^2
    (\textcolor{ucp-color}{\ell_7})^2}
  \ N\left[
    \IUCContrib{1}{4}{3}{2}{\ell_5}{\ell_6}{\ell_7}
                              \right] \notag \\
  &\quad + 
  \frac{\delta(\textcolor{cut-color}{\ell_1^2})
    \delta(\textcolor{cut-color}{\ell_2^2})
    \delta(\textcolor{cut-color}{\ell_3^2})
    \delta(\textcolor{cut-color}{\ell_4^2})}
    {(\textcolor{ucp-color}{\ell_5})^2
    (\textcolor{ucp-color}{\ell_6})^2
    (\textcolor{ucp-color}{\ell_7'})^2}
  \ N\left[
    \IUCContrib{1}{3}{4}{2}{\ell_5}{\ell_6}{\ell_7'}
    \right] + 25\text{ others.}
\end{align}
When drawing cut diagrams, we will use gray blobs as in the left-hand
sides of \cref{eq:example-cut-prod,eq:example-cut-sum} and all drawn
legs are assumed to be on-shell.  When drawing numerator contributions
to cuts, we will not use blobs on vertices, and will use
\textcolor{cut-color}{dashed lines} bisecting edges to denote cut
lines and use colored edges to highlight the
\textcolor{ucp-color}{uncut propagators}.  From now on we will leave
the cut $\delta(\ell^2)$ factors implicit.

In color-charged theories, both sides of the cut can be further
decomposed according to the color algebra.  For theories charged under
the adjoint of $SU(N)$ via $f^{abc}$-dressed amplitudes, it is
convenient to project onto a specific element of the Del
Duca-Dixon-Maltoni (DDM) color basis \cite{DixonMaltoni}.  This is done by only
inserting color-ordered amplitudes, \emph{a la} the right hand side of
\cref{eq:ddm-basis} in the product of trees, and restricting the set
of compatible diagrams to those whose color factors reduce to the
appropriate DDM element when summing over \emph{only the uncut color
  contractions}.  In this situation, the color becomes an overall
factor on the entire cut equality and thus can be ignored.  Cuts that
are organized in this manner are known as ``color-ordered'' cuts, in
the same sense as color-ordered tree amplitudes.  Since the DDM basis
(and dual Kleiss-Kuijf amplitude basis \cite{Kleiss:1988ne}) have
$(n-2)!$ elements, one generally needs to evaluate
$(n_1-2)! (n_2-2)!...$ separate color-orderings of the same cut, where
$n_1$, $n_2$... are the multiplicities of the amplitudes making up the
cut. As an example, consider the ``bowtie'' cut topology -- one of the
factorization channels of the iterated unitarity cut -- that has two
topologically distinct color-ordered expansions
\begin{equation}
  \label{eq:kt-color-ord}
\!\!\! \!\!\! \!  \NMCutC =
  \begin{cases}
     s\text{-}t \text{ channel}
    &
      \frac{1}{\textcolor{ucp-color}{s_{12}}}
      N \left[\scalebox{0.65}{\KTBTContrib{1}{4}{3}{2}}
      \right]
      +
      \frac{1}{\textcolor{ucp-color}{\ell_1^2}} \
      N \left[
      \scalebox{0.65}{\KTDBContrib{1}{4}{3}{2}}
      \right]\\
    t\text{-}u \text{ channel}
    & 
      \frac{1}{\textcolor{ucp-color}{\ell_1^2}} \
      N \left[
      \scalebox{0.65}{\KTDBContrib{1}{4}{3}{2}}
      \right]
      +
      \frac{1}{\textcolor{ucp-color}{\ell_2^2}} \
      N\left[
      \scalebox{0.65}{\KTDBContrib{1}{3}{4}{2}}
      \right]
  \end{cases}
\end{equation}
which serve as a complete basis for the full cut.

Even though unitarity methods aim to generate loop-level data from
well-defined on-shell information, the procedure can encounter
subtleties.  Beginning at one loop, there are classes of cuts that are
difficult to properly define due to hidden non-physical singularities.
The quintessential example of such a ``pathological'' cut is the 3-1
two-particle cut at one loop which hides a bubble-on-external-leg (BEL)
singularity
\begin{equation}
\frac{\delta(\textcolor{cut-color}{k_4^2})}{s\,\textcolor{ucp-color}{ k_4^2}}
n \left[
  \begin{tikzpicture}
    \ver{e1}{-1,0};
    \ver{e2}{-0.6,0.4};
    \ver{e3}{-0.3,0.4};
    \ver{i1}{-0.6,0};
    \ver{i2}{-0.3,0};
    \ver{i3}{0,0};
    \ver{i4}{0.8,0};
    \ver{co}{$(i3)!0.5!(i4)$};
    \ver{e4}{1.2,0};
    \draw[ultra thick] (e1)--(i1)--(i2)--(i3) to [out=80,in=100,min distance=0.4cm](i4)--(e4);
    \draw[ultra thick](i1)--(i2);
    \draw[ultra thick,ucp-color](i2)--(i3);
    \draw[ultra thick,cut-color](i4)--(e4);
    \draw[ultra thick] (i3) to [out=-80,in=-100,min distance=0.4cm](i4);
    \draw[ultra thick] (e2) -- (i1);
    \draw[ultra thick] (e3)--(i2);
    \draw[ultra thick,dashed,cut-color] (co)--++(90:0.5);
    \draw[ultra thick,dashed,cut-color] (co)--++(-90:0.5);
  \end{tikzpicture}
  \right]
  \in   \begin{tikzpicture}
\begin{feynman}
\vertex (a1) at (-.8, -0.6) {1};
\vertex (a2) at (-.8, 0.6) {3};
\vertex (a3) at (-1, 0) {2};
\vertex [dot, scale=2](mid1) at (0,0){};
\vertex [dot, scale=1.5,hgrey0](mid2) at (0,0){};
\vertex [dot, scale=2](mid3) at (1,0){};
\vertex [dot, scale=1.5,hgrey0](mid4) at (1,0){};
\vertex (a4) at (1.8, 0) {4};
\diagram{
(mid1) --[ultra thick,](a1),
(mid1) --[ultra thick,](a2),
(mid1) --[ultra thick,](a3),
(mid1) --[ultra thick,out=60,in=120,min distance=0.4cm](mid3),
(mid1) --[ultra thick,out=-60,in=-120,min distance=0.4cm](mid3),
(mid3) --[ultra thick](a4),
};
\end{feynman}
\end{tikzpicture}
\end{equation}
in which one of the \textcolor{ucp-color}{uncut propagators} is set
equal to an \textcolor{cut-color}{external on-shell propagator}.
Discussions about how to resolve these types of singularities can be
found in Refs.~\cite{Minahan:1987ha, Bern:2012uf, 
  Geyer:2015bja,Geyer:2015jch,Edison:2020uzf, Edison:2022jln, Edison:2022smn} and \cref{BELreg}. In addition to the
BEL cuts, there is a related class of cuts in which
an uncut propagator is set to zero by an \emph{internal cut
  condition}, for instance,
\begin{equation}
  \label{eq:other-path}
  \begin{tikzpicture}
    \pgfmathsetmacro{\ri}{1};
    \pgfmathsetmacro{\re}{0.6};
    \pgfmathsetmacro{\ang}{35};
    \begin{feynman}
      \blob{i1}{\ang:\ri};
      \blob{i2}{180-\ang:\ri};
      \blob{i3}{180+\ang:\ri};
      \blob{i4}{-\ang:\ri};
      \blob{i5}{ $(i1)!0.5!(i2)$ };
      \ver{e1}{45:\ri+\re};
      \ver{e2}{135:\ri+\re};
      \ver{e3}{225:\ri+\re};
      \ver{e4}{-45:\ri+\re};
      
      \diagram{
        {[edges={ultra thick}] (e1)--(i1),
          (e2)--(i2),
          (e3)--(i3),
          (e4)--(i4),
          (i5)--(i1)--(i4)--(i3)--(i2),
          (i2) --[out=60,in=180-60] (i5),
          (i2) --[out=-60,in=180+60] (i5)
        };
      };
    \end{feynman}
  \end{tikzpicture}
  \text{ contains a contribution from }
  \begin{tikzpicture}
    \pgfmathsetmacro{\ri}{1};
    \pgfmathsetmacro{\re}{0.6};
    \pgfmathsetmacro{\ang}{35};
    \pgfmathsetmacro{\cang}{90};
    \pgfmathsetmacro{\cl}{1.1};
      \ver{i1}{\ang:\ri};
      \ver{i2}{180-\ang:\ri};
      \ver{i3}{180+\ang:\ri};
      \ver{i4}{-\ang:\ri};
      \ver{i5}{ $(i1)!0.25!(i2)$ };
      \ver{i6}{ $(i1)!0.75!(i2)$ };
      
      \ver{e1}{45:\ri+\re};
      \ver{e2}{135:\ri+\re};
      \ver{e3}{225:\ri+\re};
      \ver{e4}{-45:\ri+\re};

      \ver{c1}{0:\cl};
      \ver{c2}{\cang:\cl};
      \ver{c3}{2*\cang:\cl};
      \ver{c4}{3*\cang:\cl};
      \ver{ci}{$(i5)!0.5!(i1)$};
      \path let \p1 =(ci) in coordinate (c5) at (\x1,0.2){};
      \draw[ultra thick] (e1)--(i1);
      \draw[ultra thick] (e2)--(i2);
      \draw[ultra thick] (e3)--(i3);
      \draw[ultra thick] (e4)--(i4);
      \draw[ultra thick] (i5)--(i1)--(i4)--(i3)--(i2)--(i6);
      \draw[ultra thick] (i6) to[out=60,in=180-60] (i5);
      \draw[ultra thick] (i6) to[out=-60,in=180+60] (i5);
      \draw[dashed,ultra thick,cut-color] (0,0) -- (c1);
      \draw[dashed,ultra thick,cut-color] (0,0) -- (c2);
      \draw[dashed,ultra thick,cut-color] (0,0) -- (c3);
      \draw[dashed,ultra thick,cut-color] (0,0) -- (c4);
      \draw[dashed,ultra thick,cut-color] (c5) -- ++(90:0.7*\cl);
      \node[fill=white,circle] (c) at (0,0) {};
  \end{tikzpicture}.
\end{equation}
Fully resolving both of these problems is beyond the scope of the
current work, so we will simply not impose any constraints that
require handling these types of cuts.

\section{One-loop cubic construction}
\label{sec:cubic}
Here we will describe why constructing the full Yang-Mills kinematic
algebra (outside the self-dual sector) is generically hard, and why
the same reason makes NLSM beyond one-loop hard as well. The major
obstacle stems from the vector state sum mixing with higher-point
contacts. Starting with the Yang-Mills Lagrangian,
\begin{equation}
  \label{eq:ym-lag}
  \mathcal{L}^{\text{YM}} = -\frac{1}{2}(\partial_\mu A_\nu)^2
  + f^{abc} \partial_\mu A^a_\nu A^b_\mu A^c_\nu
  + f^{abe}f^{ecd}A^a_\mu A^b_\nu  A^c_\mu A^d_\nu 
\end{equation}
we can express the YM three-point interaction in terms of the
following vertex in Lorenz gauge:
\begin{equation}
\Acubic{hgrey0}{}{nhpRed}{}{nhpRed}{fermion2}{} =(\varepsilon_i\varepsilon_j)(\varepsilon_k p_i).
\end{equation}
This kinematic vertex is antisymmetric in $1\leftrightarrow 2$
exchange, and from it we can construct the full Lorenz gauge
Yang-Mills vertex by summing over cyclic permutations,
\begin{equation}
V^{(123)}_{\text{YM}} = \Acubic{hgrey0}{}{nhpRed}{}{nhpRed}{fermion2}{}  +\text{cyc}(123)\,.
\end{equation}
Contracting this with the vector state sum, we can see that the cubic
Yang-Mills vertex does not satisfy the four-point Jacobi identity,
\begin{equation}
  \label{cubicJac}
  {}_1\langle V^{2}_{\text{YM}}V^{3}_{\text{YM}}\rangle_4+\text{cyc}(234) = \varepsilon_{(12)}  \varepsilon_{(34)} (s_{13}-s_{23}) +\text{cyc}(234) \neq 0
\end{equation}
where $ \varepsilon_{(ij)} = \varepsilon_{i}\cdot \varepsilon_{j}$. One way to absorb the remainder is with the four-point contact of \cref{eq:ym-lag}, which preserves non-abelian gauge invariance off-shell.
However, this additional term can also be absorbed into the definition of the
four-point kinematic numerator in a way that preserves the cubic graph construction as follows,
\begin{equation}
{}_1\langle V^{2}_{\text{YM}+B}V^{3}_{\text{YM}+B}\rangle_4 = {}_1\langle V^{2}_{\text{YM}}V^{3}_{\text{YM}}\rangle_4 + s_{12} (\varepsilon_{(13)}\varepsilon_{(24)}-\varepsilon_{(14)}\varepsilon_{(23)}).
\end{equation}
Note that this new definition of the $s$-channel numerator contains two factors of $\varepsilon_{(ij)} $ that contract spacetime indices \textit{across} the factorization channel, suggesting the inclusion of a spin-2 mode. As such, we use $V_{\text{YM}+B}$ to denote the inclusion of an additional two-form, as was studied in Ref.~\cite{Ben-Shahar:2022ixa} for constructing the NMHV Yang-Mills Lagrangian.  After the introduction of the two-form, the
numerators now satisfy the Jacobi identity
\begin{equation}
  {}_1\langle
  V^{2}_{\text{YM}+B}V^{3}_{\text{YM}+B}\rangle_4+\text{cyc}(234) =0\,,
\end{equation}
where the new cubic Lagrangian takes the form,
\begin{equation}
\mathcal{L}^{\text{YM}+B} = -\frac{1}{2}(\partial_\mu A_\nu)^2 -
B_{\mu\nu}\Box\tilde{B}_{\mu\nu}+ f^{abc} \partial_\mu A^a_\nu A^b_\mu
A^c_\nu+ f^{abc} (B_{\mu\nu}+\Box \tilde{B}_{\mu\nu})^a A^b_\mu
A^c_\nu\, .
\end{equation}
Of course, higher multiplicity would likely require further redefinition of
the three-point vertex to satisfy Jacobi identities on all internal
edges. While introducing successively higher spin states could in
principle work at tree level, we will see that it is not consistent
with what we find at general loop order. We will comment on this in
\cref{sec:Discussion}. For
now, we will study how to construct one-loop integrands consistent
with color-kinematics by isolating the cubic sector of the state
sum above.

As can be seen above, the Jacobi identity of \cref{cubicJac} fails
only in terms with two factors of polarization dot products,
$(\varepsilon\varepsilon)^2$. If we restricted ourselves to
just considering factors with a single polarization dot product, the
Jacobi identity of Yang-Mills would be satisfied \textit{off-shell}
and in arbitrary spacetime dimensions
\begin{equation}
{}_1\langle V^{2}_{\text{YM}}V^{3}_{\text{YM}}\rangle_4\big|^{(\varepsilon\varepsilon)^1}+\text{cyc}(234) =0\,.
\end{equation}
This restriction to the $(\varepsilon\varepsilon)^1$ cubic sector of Yang-Mills is closely related to the MHV decomposition of color-dual numerators \cite{TasiLance,ElvangHuangReview,Chen:2019ywi,Chen:2021chy}.
In the MHV sector it is possible to choose the reference vectors so that all dot products of polarization vectors vanish, $(\varepsilon\varepsilon) \rightarrow 0$, except those involving one of the positive helicity polarizations, $(\varepsilon^+_1 \varepsilon^-_i )\neq 0$.
We describe this in detail in \cref{Conventions}.
By focusing on $(\varepsilon\varepsilon)$ structure rather than specific 4D helicity states, we will be able to construct dimension agnostic color dual integrands at one-loop.

Thus for our approach at one-loop, we aim to build color-dual integrands directly from $D$-dimensional kinematic factors that are $\mathcal{O}((\varepsilon\varepsilon)^1)$ at tree-level, and $\mathcal{O}((\varepsilon\varepsilon)^0)$ at one-loop. The $D$-dimensional organizational principle underlying this construction can be understood in terms of a vector amplitude decomposition introduced by one of the authors \cite{Pavao:2022kog},
\begin{equation}\label{eq:pureVecRABD}
A_{(\sigma)}^{\text{YM}} = \sum_{k=0}^{\lfloor |\sigma|/2\rfloor}\sum_{\rho \in S^{2|k}_{\sigma}}\varepsilon_{(\rho)} \Delta_{(\sigma)}^{(\rho)}\,.
\end{equation}
where we have introduced the shorthand notation, $\varepsilon_{(ij)\cdots (kl)} \equiv (\varepsilon_i \varepsilon_j)\cdots (\varepsilon_k \varepsilon_l)$, and  $S^{2|k}_{\sigma}$ is the set of $k$ pairs of external legs appearing in the color-ordered label list, $\sigma$. For example, at four-point, $S^{2|1}_{\sigma} =  \{(12),(13),(14),(23),(24),(34)\}$, and $S^{2|2}_{\sigma} =  \{(12)(34),(13)(24),(14)(23)\}$. The sum thus selects out kinematic building blocks, $\Delta_{(\sigma)}^{(\rho)}$,that are each weighted by different polarization dot products. Some simple four-point examples include, 
\be
\Delta^{(13)}_{(1234)} =\frac{(k_1 \varepsilon_2)(k_{3} \varepsilon_4)}{s_{12}}+\frac{(k_1 \varepsilon_4)(k_{3} \varepsilon_2)}{s_{14}}, \qquad \Delta^{(13)(24)}_{(1234)} =1, \qquad \Delta^{(12)(34)}_{(1234)} =\frac{s_{13}}{s_{12}}\,. 
\ee
We direct the reader to Ref.~\cite{Pavao:2022kog} for further details. The expansion of \cref{eq:pureVecRABD} comes with the added advantage of making the transmutation relations of \cite{Cheung:2017ems} absolutely manifest. By a simple mass-dimension argument \cite{ElvangHuangReview}, we know the tree-level cubic sector of Yang-Mills must be at
$\mathcal{O}((\varepsilon\varepsilon)^1)$,
\begin{equation}
A_{(\text{tree})}^{\text{cubic-YM}} = \sum_{\rho \in S^{2|1}_{\sigma}}\varepsilon_{(\rho)} \Delta_{(\sigma)}^{(\rho)}\, ,
\end{equation}
while at one-loop, the cubic sector corresponds to
$\mathcal{O}((\varepsilon\varepsilon)^0)$ in polarization dot
products,
\begin{equation}
A_{(\text{1-loop})}^{\text{cubic-YM}} = \sum_{\rho \in S^{2|0}_{\sigma}}\varepsilon_{(\rho)} \Delta_{(\sigma)}^{(\rho)}\,.
\end{equation}
When decomposed in this way, the polarization stripped building blocks
must obey a set of Ward-identities between different ``helicity'', or
$(\varepsilon \varepsilon)^n$, sectors in order for the full amplitude
to be gauge invariant,
 \begin{equation}\label{eq:GIrelA}
\Delta_{(\sigma)}^{(\rho)}\Big|_{\epsilon_i\rightarrow k_i} =-
\sum_{j \in \rho^c} (k_i \varepsilon_j)\Delta_{(\sigma )}^{(\rho\cup (ij))} \,.
\end{equation}
With this in hand, we can construct a Lagrangian description of the of
the cubic sector for Yang-Mills and will demonstrate that the
resulting amplitudes with manifestly color-dual Feynman rules are
equivalent to both SDYM and NLSM through one-loop.

\subsection{Semi-abelian Yang-Mills theory}\label{semiYM}
As we argued above, if we include factors of
$(\varepsilon\varepsilon)^{n\geq 1}$ at tree-level, then we need to
keep the Yang-Mills four-point contact for color-kinematics duality to
be restored on-shell. However, the contrapositive is also true -- if
we omit the four-point Yang-Mills vertex, then we only need terms that
contribute to the manifestly color-dual cubic sector of the theory. We call
this manifestly color-dual theory \textit{semi-abelian Yang-Mills},
\begin{eBox}
\begin{equation}
\label{eq:semiYM}
\mathcal{L}^{\text{semi-YM}} = -\frac{1}{2}\text{tr}\left[\bar{F}_{\mu\nu}F^{\mu\nu}\right]
\end{equation}
\end{eBox}
where
\begin{equation}
\bar{F}^a_{\mu\nu} = \partial_\mu \bar{A}_\nu^a - \partial_\nu \bar{A}_\mu^a
\end{equation}
\begin{equation}
F^a_{\mu\nu} = \partial_\mu A_\nu^a - \partial_\nu A_\mu^a + f^{abc} A_\mu^b A_\nu^c
\end{equation}
and $A_\mu$ is in Lorenz gauge. We can construct this Lagrangian from \cref{eq:ym-lag} by keeping the right field strength covariant under
$U(N)$, and making the left field strength gauge covariant under
$U(1)^{N^2}$. Note that since $U(1)$ covariance is identical to $U(1)$ invariance, the amplitudes of this semi-abelian theory vanish under $\bar{\varepsilon}(k) \rightarrow k$, where $\bar{\varepsilon}$ is the polarization of an external $\bar{A}_\mu$ vector.

Here we can think of the abelian vector as a background field that
sources on-shell $A_\mu$ currents. Indeed, as we describe in
\cref{sec:CKLagrangians}, semi-abelian Yang-Mills theory is at the
heart of $YZ$-theory \cite{Cheung:2016prv}, $J$-theory
\cite{Cheung:2020djz,Cheung:2021zvb}, self-dual Yang-Mills
\cite{Monteiro2011pc}, and Chern-Simons theory
\cite{Ben-Shahar:2021zww}. Specifically, semi-abelian YM is simply a
clever reinterpretation of the Lagrangian obtained by integrating an
auxiliary field into the $J$-theory equations of motion.  The Feynman
rule for the cubic vertex is
\begin{equation}\label{semiYMvert}
  \cubic{hgrey0}{fermion2}{}{fermion2}{}{fermion2}{}= (\varepsilon_1\bar{\varepsilon}_3)(\varepsilon_2 k_3)
  -  (\varepsilon_2\bar{\varepsilon}_3)(\varepsilon_1 k_3)
\end{equation}
where the incoming arrow is the abelianized gauge field, and the outgoing arrows the on-shell non-abelian vectors. The propagator is simply
\be\label{semiYMprop}
\propFrule{$A^\nu$}{$\bar{A}^\mu$}{$k \rightarrow $}{fermion2} = \frac{i}{k^2} \eta^{\mu\nu}.
\ee
We can use the Ward identity of
\cref{eq:GIrelA} to check that the amplitudes are indeed gauge
invariant when the \emph{abelian} polarizations are taken to be longitudinal,
$\bar{\varepsilon} \rightarrow k$.
Furthermore, a simple calculation shows that the four-point correlation function of this theory satisfies the Jacobi identity of \cref{jacID} \emph{off-shell}. This ensures that color-kinematics duality holds to all multiplicity and loop order.
We also note that semi-YM does not have any ghosts from non-abelian gauge symmetry that could spoil color-kinematics at loop level. Due to the Feynman rules of this theory, the amplitudes are non-vanishing only at tree-level and one-loop. We demonstrate the implications of this property in \cref{2LoopObstruction}.

As noted above, we can select out the manifestly cubic
sector of semi-abelian Yang-Mills from the full theory of \cref{eq:ym-lag} by selecting only $(\varepsilon^+\varepsilon^+)$ in light-cone gauge (i.e., SDYM) and one-minus at tree-level or by plugging in the on-shell states of $J$-theory \cite{Cheung:2020djz,Cheung:2021zvb} and
$YZ$-theory \cite{Cheung:2016prv}. Indeed, semi-YM is just the following sum over building blocks
in the expansion of Yang-Mills given in Ref.~\cite{Pavao:2022kog},
\begin{equation}
A(A_1,...,\bar{A}_i,...,A_n)_{\text{tree}} = \sum_{i\neq j} (\varepsilon_i \varepsilon_j) \Delta^{(ij)}_{\text{tree}} \sim \sum_{i\neq j} (\varepsilon_i \varepsilon_j)  \sum_a \,[(\varepsilon k)^{n-2} (kk)^{3-n}]_a\,,
\end{equation}
and similarly so at one-loop,
\begin{equation}
A(A_1,...,A_n)_{\text{1-loop}} = \Delta^{(\varnothing)}_{\text{1-loop}} \sim \sum_a\,[ (\varepsilon k)^{n} (kk)^{-n}]_a\,,
\end{equation}
where $[\,\cdots]_a$ are terms with appropriate powers of
$(\varepsilon k)$ and $(kk)$. As we describe in detail in
\cref{sec:CKLagrangians}, to recover NLSM at one-loop we need to
extract the $D$-dependent part that corresponds to an internal
$\bar{Y}Y$-loop from extra-dimensional scalars
\begin{eBox}
\begin{equation}
A^{\text{NLSM}}_{\text{1-loop}} \equiv \partial_D \Delta^{(\varnothing)}_{\text{1-loop}}\big|_{\varepsilon\rightarrow k}\, .
\end{equation}
\end{eBox}
Why must we take the derivative with respect to $D$? After all,
$J$-theory and $YZ$-theory both have well defined propagators for
internal $\bar{J}J$ and $\bar{Z}Z$ propagators. However, as we discuss
in \cref{sec:CKLagrangians}, all the $J$ and $Z$ states must be on-shell in order to
produce NLSM amplitudes. Thus, the unitarity cuts of $J$ theory at
one-loop will not produce NLSM amplitudes. However, the internal
$YY$-loop is a valid forward limit for producing NLSM amplitudes,
since as constructed $YZ$-theory matches to NLSM for off-shell
$Y$-particles. We now provide explicit expressions for these one-loop amplitudes in the next section.

\subsection{One-loop color-dual integrands}\label{oneLoopCK}
Our first application of this theory for color-dual construction at
loop level is for self-dual Yang-Mills (SDYM). As we have done
throughout the text, we will set all coupling constants to unity. At
tree-level, the off-shell cubic vertices can be written in terms of
light cone coordinates \cite{Monteiro2011pc}
\begin{equation}
X(p,k) = p_u k_w-p_w k_u.
\end{equation}
A derivation of this Feynman rule and the definition of light cone coordinates can be found in \cref{sec:CKLagrangians}. At loop-level, this construction needs to be analytically continued to
general dimension in order to apply dimensional regularization at
one-loop. This can be acheived by using the cubic semi-YM vertices of the
previous section,
\begin{equation} \label{DdimSDYM}
\cubic{hgrey0}{fermion2}{}{fermion2}{}{fermion2}{}= \mathcal{X}(k_1,k_2) = (\varepsilon_1\bar{\varepsilon}_3)(\varepsilon_2 k_3) -  (\varepsilon_2\bar{\varepsilon}_3)(\varepsilon_1 k_3).
\end{equation}
Plugging in on-shell all-plus helicity states in light-cone gauge will
yield precisely the 4D SDYM vertex, up to an unphysical phase
  \begin{equation}
  \mathcal{X}(k^+_1,k^+_2) = \frac{\langle 12\rangle^3}{ \langle23\rangle \langle 31\rangle} \sim \langle 12\rangle = k_{1,u} k_{2,w}-k_{1,w} k_{2,u}\, ,
 \end{equation}
where $\mathcal{X}(k^+_1,k^+_2) = A^{\text{YM}}(1^+,2^+,3^-)$ in light cone gauge.
In the second equality, momentum conservation has been applied to the redefined the spinor bracket, $\langle 12\rangle \rightarrow X(k_1,k_2)$, whose definition is given in \cref{Conventions}.
Of course, the form of \cref{DdimSDYM} has the advantage of
permitting a $D$-dimensional construction of the one-loop
integrand. As an example, the four-point box numerator is
\begin{equation}
\simplebox = \langle \mathcal{X}(k_1,\ell_1)\mathcal{X}(k_2,\ell_2)\mathcal{X}(k_3,\ell_3)\mathcal{X}(k_4,\ell_4)\rangle \, ,
\end{equation}
where $\ell_i = \ell-(k_1+k_2+\cdots+ k_i)$ and the bracket $\langle \,\cdots\rangle$ indicates that we have applied the gauge fixed state projector, $\sum {\varepsilon^\mu_{(+\ell_i)}\varepsilon^\nu_{(-\ell_i)}}=\eta^{\mu\nu}$, on all internal polarizations.
In general, the $n$-gon diagram is
\begin{equation}\label{SDYMnGon}
N^{\text{SDYM}}_{n\text{-gon}}= \nGonVector = \langle \mathcal{X}(k_1,\ell_1)\mathcal{X}(k_2,\ell_2)\cdots \mathcal{X}(k_n,\ell_n) \rangle\, .
\end{equation}
Recall that all other numerators can be obtained from the $n$-gon through Jacobi.
Due to the state-sum of internal loop factors,
$\sum {\varepsilon_{(+\ell)}\varepsilon_{(-\ell)}}\sim D$, the integrand above
depends explicitly on the spacetime dimension, $D$. By taking a
derivative\footnote{Another way to understand this derivative is that
  it selects the large $D$ behavior of the integrand.}, we can recover
the integrand needed for the all-plus one-loop amplitudes with an internal
scalar
\begin{equation}
\partial_D N^{\text{SDYM}}_{n\text{-gon}}=\nGonScalar  = 2^n (\ell_1 \varepsilon_1)(\ell_2 \varepsilon_2)\cdots (\ell_n \varepsilon_n)\, .
\end{equation}
This integrand numerator with internal scalar loop is precisely what
one would obtain from the Feynman rules of $YZ$-theory, absent the $\bar{Z}Z$
internal vector loop. For more background, we refer the reader to
\cref{sec:CKLagrangians}. When plugging on the on-shell states of $YZ$-theory, we thus obtain the following expression for the NLSM one-loop $n$-gon numerator
\begin{equation}\label{NLSMnGon}
N^{\text{NLSM}}_{n\text{-gon}}=\left[\partial_D  N^{\text{SDYM}}_{n\text{-gon}}\right]^{\epsilon \rightarrow k}=\nGonScalar =   [\![ 12]\!][\![23]\!]\cdots [\![n1]\!]\, ,
\end{equation}
where we have defined the antisymmetric kinematic variable,
$[\![ij]\!] = \ell_i^2 - \ell_j^2 = 2(k_i \cdot \ell_i)$.
We have
verified through 10-point one-loop that this $n$-gon expression is a
valid color-dual representation for NLSM. Thus, composing the $n$-gon numerators of \cref{SDYMnGon} and \cref{NLSMnGon} would yield $D$-dimensional integrands that project down to the 4D all-plus Born-Infeld one-loop amplitudes studied in Ref.~\cite{Elvang:2020kuj}.

Before proceeding, we note that the above definition does give rise to ``pathological" bubble-on-external-leg (BEL) diagrams, discussed previously in \cref{sec:bootstrap}. However, one can show that these diagrams integrate to zero for spacetime dimension, $D>2$, and thus can be disregarded as unphysical. For the interested reader, in \cref{BELreg} we provide a detailed overview of this dimensional regularization of the relevant BEL diagram.
\subsection{Two-loop obstruction}\label{2LoopObstruction}
At two-loop, introducing terms that conspire with four-point contacts
is unavoidable. At one-loop, we were able to avoid internal
contractions of $\bar{A}_\mu A^\mu$ by selecting appropriate external
states. However, at two-loop when choosing all external $A_\mu$
states, the amplitude vanishes in semi-YM theory
\begin{equation}
\mathcal{A}^{\text{semi-YM}}_{\text{2-loop}}(A_\mu,A_\mu,A_\mu,A_\mu)=0,
\end{equation}
that is, the theory is one-loop exact.  In order to produce
non-vanishing interactions, we would need to reintroduce
$D$-dimensional vertices from the full Yang-Mills Lagrangian of
\cref{eq:ym-lag} that we dropped in our construction semi-abelian
YM. In terms of cubic graphs, the additional interaction must
necessarily have the opposite number of $\bar{A}_\mu$ and unbarred
$A_\mu$ fields to that of \cref{semiYMvert}.  Reintroducing these
oppositely oriented vertices allows for new internal contractions of
$\bar{A}_\mu A^\mu$,
\begin{equation}\label{2loopTension}
\mathcal{A}^{\text{YM}}_{\text{2-loop}}(A_\mu,A_\mu,A_\mu,A_\mu)=\doubleBoxObstructionA +\doubleBoxObstructionB+\cdots 
\end{equation}
where we used white dots to indicate interaction vertices of
weight\footnote{We choose the convention that the abelianized gauge
  fields carry weight $\mathcal{W}[\bar{A}] = -1$, and non-abelian
  vectors are $\mathcal{W}[\bar{A}] = +1$.}
$\mathcal{W}[\bar{A}\bar{A} A] = -1$, rather than the isolated semi-YM
vertex which always carries weight $\mathcal{W}[\bar{A}A A] =
+1$. This immediately runs into the difficulty of introducing the
four-point contact needed for color-kinematics to be satisfied on all
internal edges. Therefore, by a simple weight counting argument, one can see that
including this wrong sign interactions is unavoidable at two-loop and
higher.

Thus, to construct two-loop numerators prescriptively, as we have done at one-loop, would require knowledge of
the full kinematic algebra for Yang-Mills off-shell.  Since this is presently unavailable, we will tackle the two-loop integrand using an ansatz approach.

\section{Two-loop four-point bootstrap}
\label{2loopBoot}
As we have just seen, it is possible to coerce tree numerators into
one-loop numerators at any multiplicity for pions and related
theories, but that these methods cannot be reapplied to generate
higher-loop numerators.  For pions in particular, the two-loop no-go statement
is only for \emph{one particular representation of the theory}, so it does
not completely preclude the existence of a two-loop color-dual
integrand.  We thus turn to the color-dual bootstrap method
described in \cref{sec:bootstrap} to construct a color-dual
representation of two-loop four-point NLSM.  Because of the similarity
of the problem setup, we will also use the opportunity to revisit the
construction of a color-dual representation of pure YM, extending the
search space beyond what was covered in Ref.~\cite{Bern:2015ooa} to include the most general local ansatz.

Both NLSM and YM share the same cubic graph basis, and thus their
defining Jacobi relations lead to the same graph basis.
Ignoring tadpole and BEL graphs, there are 14 cubic four-point two-loop
graphs (see \cref{fig:MaxCuts}) which are related to each other via 21
Jacobi relations.  As mentioned in \cref{sec:bootstrap},
color-kinematics duality ensures that the numerator of every graph can
be expressed in terms of a basis of the double box and penta-triangle,
\begin{equation}
\label{eq:jacobi-basis}
\NLSMBasisDoubleBox \text{ and } \NLSMBasisPentaTriangle .
\end{equation}
Obviously, the actual numerator dressings and physical properties
of the two theories differ, so we discuss them separately below.

\subsection{Two-loop NLSM}
\label{sec:pions}

For a scalar theory like NLSM the numerator only depends on dot
products of momenta.  The basis of momentum invariants for both of the
graphs can be taken to be
\begin{equation}
\label{eq:scalar-basis}
N^{\text{NLSM}}\left[
  \scalebox{0.75}{\NLSMBasisDoubleBox} \right]
\text{ and }
N^{\text{NLSM}}\left[\scalebox{0.75}{\NLSMBasisPentaTriangle} \right]\in
\left\{\begin{array}{cccc}
    k_{13},& k_{23},& k_{15},& k_{16}, \\
    k_{25},& k_{26},& k_{35},& k_{36}, \\
    k_{55},& k_{56},& k_{66}
  \end{array}
  \right\},
\end{equation}
where $k_{ij}$ means $k_i \cdot k_j$, $k_5=\ell_1$, and $k_6 = \ell_2$.
Every vertex in NLSM scales as $k^2$ so the numerator of each diagram
scales as $k^{12}$.  At this mass dimension there are 8,008
independent dot products of momenta.  With two basis graphs there is a
total of 16,016 ansatz parameters.\footnote{\texttt{Mathematica} is
  not well suited for solving large sets of equations with sixteen thousand
  parameters, so a different method is desirable.  One possibility is
  to use a sparse solver, as often employed for integral reduction.
  However, these solvers typically assume that the matrix has a very
  low density.  While the equations in YM naturally separate into
  different ``helicity'' sectors (or $e_i \cdot e_j$ and
  $e_i \cdot k_j$ sectors in $D$ dimensions) which limits the
  cross-talk between equations and thus the effective density, there
  is no such separation of sectors in a scalar theory.  Thus the NLSM
  constraint equations were solved using a custom solver, working name
  \texttt{FiniteFieldSolve}, designed to exactly solve large linear
  systems of arbitrary density.  \texttt{FiniteFieldSolve} will be
  released publicly shortly.} Imposing the boundary Jacobi relations
eliminates all but 4,473 of the original free parameters.  Further
imposing graph symmetries reduces the number of free parameters to
1,243.

\begin{figure}[t]
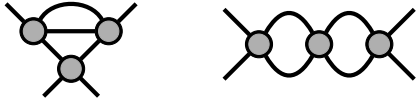

  \begin{center}
	\PhysicalCutOne{}{}{}{} \PhysicalCutTwo{}{}{}{}
  \end{center}
  \caption{The two physical cut topologies shared by NLSM, sGal, BI, and DBIVA}
  \label{fig:emu}
\end{figure}

The final essential constraint is unitarity.  The two physical cuts
that need to be enforced, shown in \cref{fig:emu}, only involve
four-point vertices.  Notably this means that even at two loops,
unitarity is still only probing the on-shell four-point NLSM amplitude.
Unitarity cuts besides those shown in \cref{fig:emu} are either
pathological or vanish because they involve a three-point point on-shell
amplitude. In order to manifest the $\mathbb{Z}_2$
symmetry of NLSM at the integrand level, we also require that all
non-pathological cuts containing a three-point vertex (such as the maximal
cut) vanish.  Imposing all relevant cuts leaves 765 free parameters
representing the generalized gauge freedom in the answer.

At this point color-kinematics duality has been satisfied, so the
double copy will proceed without issue.  However, the plethora of
generalized gauge freedom parameters leaves the option for enforcing
additional aesthetic constraints.  All remaining generalized gauge
parameters could be set to zero, but a simpler and more insightful
result can be obtained through physical arguments.  $\mathcal{N}=4$
SYM provides several hints for further conditions to impose on the
ansatz.  For example, for maximally supersymmetric gauge theory it is
possible to enforce the no triangle hypothesis, manifest loop power
counting, and, for four-point and up to at least six loops, strip off
a factor of $st \atree$ from the integrand \cite{FiveLoopN4,
  Bern:2012uf, Bern:2010fy, SuperSum, Carrasco:2021otn,
  JJHenrikReview, BRY, BDDPR, Neq44np}.  For the NLSM integrand it is
desirable to make color-kinematics duality as manifest as possible.
One hope would be to factor out some piece of the one-loop numerator
since this at least manifests antisymmetry for the vertices involving
external legs.  However, a more fruitful direction is to match onto
the one known theory with pion power counting that manifests
color-kinematics duality to all loop orders, namely,
Zakharov-Mikhailov (ZM) theory \cite{Zakharov:1973pp, Cheung:2022mix}.
ZM theory is governed by the Lagrangian
\begin{equation}
\label{eq:ZMLagrangian}
\mathcal{L}^{\text{ZM}} = \frac{1}{2}(\partial \varphi)^2
+ g f^{abc} \varphi^a \varepsilon^{\mu\nu}(\partial_\mu \varphi^b)( \partial_\nu \varphi^c),
\end{equation}
where more details can be found in \cref{sec:CKLagrangians}.  The
color-stripped Feynman rule for the vertex is
$V(k_1, k_2, k_3) \propto \varepsilon_{\mu\nu}k_1^\mu k_2^\nu \equiv \langle k_1 k_2 \rangle$ where
off-shell color-kinematics duality to all orders in perturbation
theory simply follows from the Schouten identity in 2D.  Since the theory
is purely cubic and manifests off-shell color-kinematics
duality, it is trivial to read off the color-dual numerator for any
graph.  From the presence of the Levi-Civita tensor
$\varepsilon^{\mu\nu}$, the theory clearly resides in two spacetime
dimensions where scattering is notoriously plagued by infrared
regulation issues.  Every on-shell particle is either a right mover,
with momentum proportional to $k_R^\mu \equiv (1,1)$, or a left mover,
with momentum proportional to $k_L^\mu \equiv (1,-1)$.  On-shell ZM
amplitudes naturally divide into sectors corresponding to the
configuration of left and right movers, where the scattering in many
sectors is rather subtle.  Only color-ordered amplitudes can be
defined unambiguously from the naive Feynman rules.  At four-point, only the
alternating sector is free of subtleties and the amplitude for this
process vanishes.  In equations, $A[LRLR]=0$ where $R$ corresponds to
a right mover and $L$ corresponds to a left mover.  In every other
configuration of right and left movers, such as $A[RRRR]$, one of the
internal propagators is accidentally on-shell since $k_R^2=k_L^2=0$.

It is tempting to try to match the pion four-point two-loop integrand to ZM
on the cuts, but given that the cuts only probe the on-shell four-point
amplitude, which is either subtle or vanishes for ZM, it is better to
compare the off-shell numerators directly.
The numerators of the two basis graphs are
\begin{align}
N^{\text{ZM}}\left[ \scalebox{0.75}{\ZMBasisDoubleBox} \right] &=
\begin{aligned}
&\langle \ell_1 \ell_2 \rangle \langle k_4 \ell_1 \rangle \langle \ell_2 k_1 \rangle \langle k_3 (\ell_1+k_4)\rangle \\
& \times \langle (k_1+\ell_2) k_2 \rangle \langle (\ell_2+k_{12})(\ell_1-k_{12})\rangle
\end{aligned} \\
N^{\text{ZM}}\left[ \scalebox{0.75}{\ZMBasisPentaTriangle} \right] &=
\begin{aligned}
&\langle \ell_1 \ell_2 \rangle \langle k_4 \ell_1 \rangle \langle \ell_2 k_1 \rangle \langle k_3 (\ell_1+k_4) \rangle \\
& \times  \langle (k_1-\ell_1) k_2 \rangle \langle (\ell_2+k_1) (\ell_1-k_1)\rangle
\end{aligned}
\end{align}
where $k_{12} = k_1+k_2$ and, again, $\langle ab\rangle \equiv p_a^\mu \varepsilon_{\mu\nu} p_b^\nu$ is the color-stripped ZM vertex.
As local functions, the
numerators never suffer from any of the subtleties of on-shell 2D
kinematics.
We introduce a proportionality constant $z$ between the numerators of the two theories, $N_\text{NLSM} \vert_{2D} = z ~ N_\text{ZM}$, because we are only interested in how the kinematical structure of $N_\text{ZM}$ can be used to eliminate gauge freedom in $N_\text{NSLM}$.\footnote{Another reason for introducing the parameter $z$ is because the ZM and NLSM are believed to differ at loop level even though they are dual classically in 2D \cite{Nappi:1979ig}.  Once all of the constraints from this section are imposed, $z$ is fixed to $216/565$ where the four-point pion tree amplitude is normalized to $-k_1\cdot k_2$ and the ZM vertex is normalized to $k_1^\mu \varepsilon_{\mu\nu}k_2^\nu$.}
Mechanically, the pion numerator is matched
to ZM by first taking every possible assignment of right and left
movers for the external particles and then restricting the pion
numerator to 2D.  The loop momenta are restricted to 2D but left
off-shell.  After performing the 2D matching, there are only 365
parameters of generalized gauge freedom.

The loop momentum structure of ZM theory provides one final hint for
simplifying the pion numerators.  A generic term in the ZM numerator
(of either basis graph) looks schematically like
\begin{equation}
\label{eq:ZMLoopPowCount}
\ell_1^m \ell_2^n k^{12-(m+n)} \text{ where } 5 \leq m+n \leq 8,
\end{equation}
whereas a generic set of local, cubic Feynman rules could have
produced terms with up to $m+n=12$ powers of loop momenta.  When the
pion numerators are forced to have the loop power counting structure in
\cref{eq:ZMLoopPowCount} of ZM theory, the number of generalized gauge
freedom parameters reduces to 58.  The pion numerators appearing in the
ancillary files make exactly this choice.

\subsection{Double-copy verification}
\label{doubleCopyVerify}

With a cubic color-dual pion representation in hand, we can perform
double copies with many theories to extract colorless gravity-like
amplitudes.  In particular, we produce numerators for special
Galileons (sGal), Born-Infeld (BI) theory, and Dirac-Born-Infeld-Volkov-Akulov (DBIVA) by double-copying against
pions, pure YM, and $\mathcal{N}=4$ super-Yang-Mills (sYM) respectively
\cite{Cachazo:2014xea,Cheung:2015ota,Cheung:2016drk}.  Such double-copy constructions are important nontrivial
checks on the underlying single-copy theories as the Jacobi relations
in the single copy conspire to produce the double-copy theory's
version of linear diffeomorphism invariance: enhanced shift symmetry for special Galileons and gauge
invariance of the BI photon and DBIVA supermultiplet \cite{Hinterbichler:2015pqa}.  All
three of these double-copy theories were recently studied extensively
by one of the authors and Carrasco from the perspective of direct
unitarity cut construction \cite{Carrasco:2023qgz}.  One of the shared
features of these three theories is that they all have the same
physical cut topologies in 4D: the two diagrams shown in
\cref{fig:emu}, which are only composed of four-point amplitudes.

To construct the double-copy theories, we source the cubic
pure YM numerators from Ref.~\cite{Bern:2015ooa} (which
additionally satisfies a relaxed form of color-kinematics duality),
and the cubic $\mathcal{N}=4$ sYM representation from the well-known
$n_{2\text{box}} = n_{\text{cross-box}} = s^2t\, \atree$ \cite{Bern:1997nh}.
On the other hand, we use the methods of Ref.~\cite{Carrasco:2023qgz}
to directly compute the needed basis cuts in all three theories without
relying on a color-dual pion representation.  We find exact agreement
in each of the three theories for both physical cuts.  Since
Ref.~\cite{Carrasco:2023qgz} has already exhaustively explored the
properties of four-point loop amplitudes in these theories, we direct
interested readers there for more information.

\subsection{Two-loop Yang-Mills revisited}\label{2loopYM}

\begin{figure}
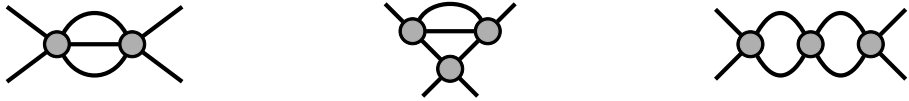

  \begin{center}
    \begin{subfigure}{0.3\textwidth}
      \begin{center}
        \LMCut
      \end{center}
    \end{subfigure}
    \begin{subfigure}{0.3\textwidth}
      \begin{center}
        \PhysicalCutOne{}{}{}{}
      \end{center}
    \end{subfigure}
    \begin{subfigure}{0.3\textwidth}
      \begin{center}
        \PhysicalCutTwo{}{}{}{}
      \end{center}
    \end{subfigure}
  \end{center}
  \caption{The three spanning physical unitarity cuts of pure
    Yang-Mills at two loops}
  \label{fig:ym-spanning}
\end{figure}

Given the close ties between pion and gluon scattering for trees and
at one loop, the existence of the two-loop pion numerator prompts us
to investigate the most general local numerator for YM, without any of
the loop power counting assumptions of Ref.~\cite{Bern:2015ooa}.  We
follow the same general procedure as in \cref{sec:pions}, again
identifying the double-box and penta-triangle as the
Jacobi basis graphs and building the most general ansatz for each of
their numerators compatible with the assumptions in
\cref{sec:bootstrap}.  Because we are now considering pure
Yang-Mills, each monomial in the local ansatz must consist of five
Lorentz scalar dot products instead of the six for NLSM, and each
monomial must be linear in each of the four external gluon
polarizations.  Thus, the numerator ansatz for both diagrams will be
built from terms of the form
\begin{equation}
  N^{\text{YM}}\left[\scalebox{0.75}{\NLSMBasisDoubleBox} \right]
  \text{ and }
  N^{\text{YM}}\left[\scalebox{0.75}{\NLSMBasisPentaTriangle} \right]
  \in \text{span}\left\{(\varepsilon_i \varepsilon_j) , (k_i \varepsilon_j), (k_ik_j)\right\}
  \label{eq:ym-basis}
\end{equation}
where the $k_i$ take the same definition as described near
\cref{eq:scalar-basis}.  Without any power counting restrictions
imposed, both the double-box and the penta-triangle have an ansatz
with 10,010 terms each.  Imposing diagram automorphisms and
maximal-cut gauge invariance on the two basis diagrams, we reduce the
number of terms to 2,235 for the double-box, and 4,133 for the
penta-triangle.  The minimal set of spanning physical cuts is shown in
\cref{fig:ym-spanning}, but we choose to work within the framework of
the method of maximal cuts \cite{Bern:2007ct} in order to identify the
simplest cut that is in tension with the kinematic Jacobi relations.
Maximal cuts and symmetries are then imposed on the 14
non-pathological cubic diagrams shown above in \cref{fig:MaxCuts}.
Doing so, we are left with 596 total parameters in the ansatz.

\begin{figure}
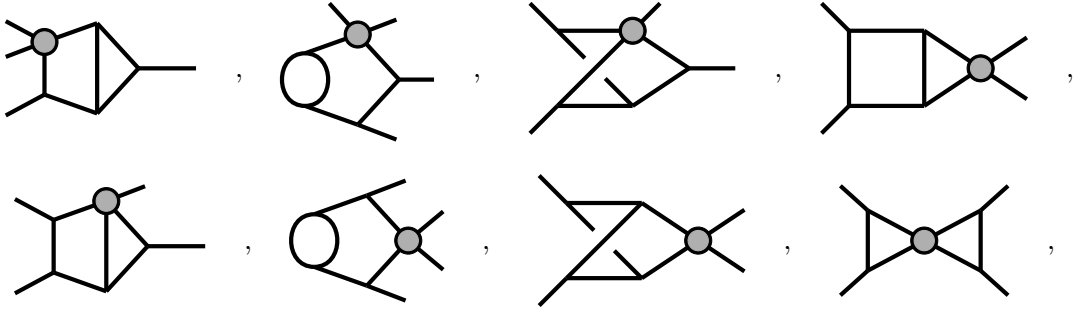

\centering
 { \NMCutH, \NMCutG,  \NMCutD,\NMCutB, 
 \\
 \NMCutE, \NMCutF, \NMCutA, \NMCutC,  }
  \caption{The 8 non-pathological 1-particle-irreducible next-to-maximal-cut diagrams used in our cut constraints on the Yang-Mills integrand.}
  \label{fig:ym-nmc}
\end{figure}

Proceeding to the next-to-maximal cuts, we continue to avoid
pathological diagrams including the newly-appearing type discussed in
\cref{eq:other-path}.  After discarding these types of cuts as well as
those involving cuts of external Mandelstams, there are 8 one-particle-irreducible cut
topologies (see \cref{fig:ym-nmc}). Seven of these next-to-maximal
cuts are consistent with the kinematic Jacobi relations and
symmetries.  Critically, the ``bowtie'' cut,
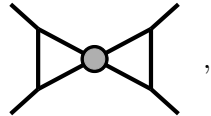
\begin{equation*}
   \NMCutC \,,
\end{equation*}
cannot be satisfied by the ansatz once Jacobi relations and
symmetries are applied.

In fact, we can make the failure extremely precise. First, start with the
diagrams shown in the Jacobi relation from \cref{eq:TwoLoopJacobiOne}, which includes
the double-box, crossed-box, and penta-triangle. Then \emph{without} defining the crossed-box in terms of the other two diagrams, write down the most generic parity-even local ansatz
involving four polarizations and six momenta (any combination of
external or loop) for each of the three diagrams.  Each ansatz will have 10,010 terms initially.  Next
impose the symmetry constraints on each of the three diagrams, leaving
2,761 free parameters on the double-box, 2,576 free parameters on the
crossed-box, and 5,040 free parameters on the penta-triangle.  Finally
impose the ``bowtie'' next-to-maximal cut, which in the
\emph{non-planar} $t$-$u$ color channel only receives contributions
from the double-box
\begin{equation}
   \NMCutC
  =
  \frac{1}{\textcolor{ucp-color}{\ell_1^2}} \
  N \left[
    \KTDBContrib{1}{2}{3}{4}
  \right]
  +
  \frac{1}{\textcolor{ucp-color}{\ell_2^2}} \
  N\left[\KTDBContrib{1}{3}{2}{4}\right] \,.
\end{equation}
After imposing this cut, we find that the original Jacobi relation,
\cref{eq:TwoLoopJacobiOne}, can no longer be satisfied. In other words, the constraints imposed by the single Jacobi relation and the three sets of diagram symmetries conspire in a way that is inconsistent with the ``bowtie'' cut. The contrapositive to this surprising result can be summarized in the following diagrammatic statement, which prioritizes the physical unitarity cut:
\begin{eBox}
\begin{equation}\label{chefsKiss}
   \scalebox{0.9}{\NMCutC} \!\!
  \Rightarrow \!\!
  \scalebox{0.85}{\JacobiOneDoubleBox}
  +  \scalebox{0.85}{\JacobiOneCrossedBox}
  + \scalebox{0.85}{\JacobiOnePentaTriangle} \neq 0 \,.
\end{equation}
\end{eBox}
Thus we have identified the minimal failure state for a
globally-color-dual two-loop Yang-Mills representation: off-shell locality,
symmetry, and the kinematic Jacobi relations for just three diagrams
are incompatible with the ``bowtie'' cut.

\section{Conclusions and Outlook}\label{conclusions}

In summary, we have found that there is a tension between color-kinematics duality and an off-shell local construction of the two-loop four-point pure Yang-Mills integrand in $D$-dimensions.
At tree level, we found that the duality can be
manifested by breaking manifest Bose symmetry at the Lagrangian level,
either by explicitly picking different gluon states (as in SDYM) or by
picking states that only restore Bose symmetry in the final amplitude
(as in NLSM).  Both of these theories can be encapsulated in the
semi-abelian YM theory presented in \cref{eq:semiYM} which obscures
the Bose symmetry of the external states due to the presence of the
two gauge fields $A_\mu$ and $\bar{A}_\mu$.  

In addition to the tree-level construction, we found that kinetic
mixing between the two gauge fields permits globally color-dual
one-loop integrands. After plugging in suitable on-shell states, the
integrands produced by semi-abelian YM correspond to those of SDYM in
\cref{SDYMnGon} and NLSM in \cref{NLSMnGon}, along with Chern-Simons
theory, which we discuss in \cref{sec:CSTheory}.  Due to the kinetic
mixing that stems from the broken Bose symmetry, \cref{2loopTension}
shows why it is impossible to construct a two-loop integrand for
semi-abelian YM theory. In \cref{2loopBoot}, we overcame this
obstruction by successfully constructing two-loop NLSM numerators with
an ansatz-based bootstrap. However, we found that the same approach
fails for two-loop pure Yang-Mills, and pin-pointed the failure to a
particular unitarity cut captured by \cref{chefsKiss}. Thus, obtaining
a globally color-dual integrand beyond one-loop then requires
sacrificing more than just Bose symmetry, and in this paper we have
argued that off-shell locality\footnote{An example of this can already
  be found in the literature, where locality is relaxed in
  Ref.~\cite{Mogull:2015adi} in order to obtain color-dual two-loop
  integrands with on-shell 4D states.} must be abandoned for
color-dual constructions of multi-loop Yang-Mills. To expand the
available function space, we propose that one should consider
including rational functions of kinematics, rather than just
polynomials.

We concede building loop-level numerators from rational functions of the kinematics
is rather unnatural from the perspective of point-like quantum field
theories. After all, operators that produce rational functions are typically non-local in their construction. The archetypal example of non-local quantum operators are string vertex operators with $\alpha'$ corrections. These operators promote local tree-level amplitudes to stringy extended objects by integrating over disc integrals of the open string worldsheet
\be
\mathcal{A}^{\text{YM}} \stackrel{\alpha'}{\longrightarrow} \mathcal{A}^{\text{OS}}\supset t_8 F^4 \int_0^1 dx \,\frac{ x^{\alpha' s_{12}}(1-x)^{\alpha' s_{23}}}{x(1-x)(s_{12}+s_{23})}\, .
\ee
The resulting Veneziano factor of Gamma functions produces rational functions of kinematics, while preserving color-kinematics duality \cite{MafraBCJAmplString,Broedel:2013tta,Mafra:2016mcc,Carrasco:2019yyn}. While introducing some type of worldsheet formulation of color-dual numerators might seem unjustified given the results of this work, our findings certainly suggest that we must do \textit{something} to relax the constraint of off-shell locality, and realize locality only in the limit of
\textit{on-shell} kinematics. This could be achieved by either modifying the kinetic term or potentially something new and more exotic. Below we provide some simple examples
at tree-level of what one might consider for implementing such a construction.

\subsection{Non-local construction of scattering amplitudes}
\label{nonLocalScattering}
As an exemplar of an off-shell non-local structure, consider the simple four-point example of a color-dual
representation of Yang-Mills theory. We can define a functional numerator, $N_{(12|34)}^{\text{YM}}$, as follows, 
\begin{equation}
N_{(12|34)}^{\text{YM}} = \frac{t_8F^4}{3} \frac{s_{13}-s_{12}}{s_{23}{s_{13}}}
\end{equation}
where $N_s = N_{(12|34)}$, $N_t = N_{(14|23)}$ and $N_u = N_{(13|42)}$. By construction, this functionally symmetric numerator is
antisymmetric and obeys the Jacobi identity
\begin{equation}\label{nonLocalNum}
N_{(12|34)}^{\text{YM}} + N_{(13|42)}^{\text{YM}} +N_{(14|23)}^{\text{YM}} = 0.
\end{equation}
However, it must also factorize to kinematics that are consistent with
the local Feynman rules of \cref{eq:ym-lag}. Evaluating the residue on
the $s=s_{12} \to 0$ pole we find
\begin{align}
\text{Res}\left[\frac{ N_{s}^{\text{YM}} }{s}\right]^{s=0}&= \frac{2}{3}A_{(12|l)}^{\text{YM}}A_{(-l|34)}^{\text{YM}} 
\\
\text{Res}\left[\frac{N_{t}^{\text{YM}} }{t}\right]^{s=0}&= -\frac{1}{3}A_{(12|l)}^{\text{YM}}A_{(-l|34)}^{\text{YM}} 
\\
\text{Res}\left[\frac{N_{u}^{\text{YM}} }{u}\right]^{s=0}&= -\frac{1}{3} A_{(12|l)}^{\text{YM}}A_{(-l|34)}^{\text{YM}} \, ,
\end{align}
with $t=s_{23}$ and $u=s_{13}$.
Plugging these into a cubic graph representation of Yang-Mills, the
tree level version of \cref{eq:gen-amp}, we find
\begin{equation}
  \text{Res}\left[\mathcal{A}_4^{\text{YM}}\right]^{s=0}
  = \frac{1}{3} A_{(12|l)}^{\text{YM}}A_{(-l|34)}^{\text{YM}} (2c_s-c_t-c_u)
  =  \mathcal{A}_{(12|l)}^{\text{YM}}\mathcal{A}_{(-l|34)}^{\text{YM}} \, ,
\end{equation}
where we have applied the color structure Jacobi identity,
$c_s+c_t+c_u=0$. In light of our findings in \cref{chefsKiss}, building color-dual numerators in this way where locality is only realized on-shell might be a more natural approach. While we can absolutely apply a generalized gauge
transformation \cite{BCJ} to restore field theoretic locality to the
cubic numerators, this could merely be an aesthetic choice that is only permissible at tree-level.

Moreover, by abandoning traditional notions of off-shell
locality, we incidentally have gained enough functional freedom to massage the
color-dual numerators into a form that is manifestly gauge invariant
for all particles. As an organizational principle, pulling out overall
gauge-invariant factors comes with the added advantage of possibly simplifying
a loop-level construction of color-dual numerators for Yang-Mills.

\subsection{Future Directions}\label{sec:Discussion}
With an eye towards generalizing the non-local construction above to future multi-loop studies, we note that the four-point half-ladder of Yang-Mills secretly makes use of the color-dual structure of NLSM. At four-point, we can construct permutation invariants from the single BCJ basis amplitude of both NLSM and pure YM theory as follows
\be
s t A^{\text{YM}}_{(s,t)} = t_8F^4 \qquad s t A^{\text{NLSM}}_{(s,t)} = s t u\, .
\ee
Using this, we can redefine the non-local numerators of \cref{nonLocalNum} so that the vector structure of YM is captured in a permutation invariant prefactor and the kinematic Jacobi identity is entirely due to the NLSM numerators
\be
N_{(12|34)}^{\text{YM}} \equiv t_8F^4 \,\frac{N_{(12|34)}^{\text{NLSM}}}{s t u } \, .
\ee
where four-point pion numerator, $N_s^{\text{NLSM}}$, is given in \cref{pionNums}. One advantage of this construction is that it reduces some of the $D$-dimensional complexity of vector theories that arises due to the mixing between external polarization and
internal loop momenta. But maybe more importantly, it puts all the heavy lifting of functional Jacobi relations on the scalar kinematic numerator, $N_{(12|34)}^{\text{NLSM}}$. Thus, rather
than building an ansatz from the irreducible scalar products of
\cref{eq:ym-basis}, one might instead consider the following construction
\begin{equation}
N^{\text{YM}}_{m,L} = \sum_i \mathcal{O}_i \mathcal{R}^{(i)}_{m,L}
\end{equation}
where $\mathcal{O}_i$ are the on-shell gauge invariant tensor basis elements of
\cite{Bern:2017tuc,Carrasco:2019yyn}, and $ \mathcal{R}^{(i)}_{m,L}$ are
rational functions of irreducible scalar products of \cref{eq:scalar-basis}. It would be interesting if a similar construction as our tree-level example could be uplifted to two-loop using the globally color-dual NLSM integrand that we have computed in this work. We see this as a natural future direction worth investigating that is now made possible by our findings.

\subsection*{Acknowledgments}
The authors would like to thank John Joseph Carrasco, Sasank Chava, Clifford Cheung, Kezhu Guo, Nia Robles, Aslan Seifi, Fei Teng, and Suna Zekio\u{g}lu for insightful conversations,
feedback on earlier drafts, and encouragement throughout the completion of
this work. This work was supported by the DOE under contract
DE-SC0015910 and by the Alfred P. Sloan
Foundation. Additionally we would like to acknowledge the Northwestern
University Amplitudes and Insight group, the Department of Physics and
Astronomy, and Weinberg College for their generous support.  Feynman
diagrams were typeset using TikZ-Feynman \cite{Ellis:2016jkw}.

\appendix
\addtocontents{toc}{\protect\setcounter{tocdepth}{1}}
\section{Catalog of color-dual theories}
\label{sec:CKLagrangians}

Below we review the five manifestly color-dual theories that we
mention in the text, self-dual Yang-Mills (SDYM)
\cite{Monteiro2011pc}, two formulations of NLSM
\cite{Cheung:2016prv,Cheung:2020djz,Cheung:2021zvb}, ZM
\cite{Zakharov:1973pp, Cheung:2022mix}, and Chern-Simons theory
\cite{Ben-Shahar:2021zww}.  All of these theories are closely tied
together by the semi-abelian YM theory in \cref{eq:semiYM}, which
gives the cubic sector of pure YM.  The manifestly color-dual
equations of motion for SDYM and NLSM are derived by picking a gauge
and then placing a constraint directly on the field strength. In
almost all of these theories, Bose symmetry is broken or obscured at
the Lagrangian level.  Since Bose symmetry is broken, the ensuing
kinetic mixing prevents color-dual representations beyond one-loop
without the introduction of additional interactions.

\subsection{Self-Dual Yang-Mills} The first example in the literature
of color-dual Feynman rules is that of Yang-Mills theory in the
self-dual sector (SDYM).  We summarize the approach of
Ref.~\cite{Monteiro2011pc} for extracting the kinematic algebra.
To obtain the Lagrangian for self-dual Yang-Mills from standard Yang-Mills, we apply the
self-duality condition that equates the non-abelian $SU(N)$ field
strength,
$F_{\mu\nu} = \partial_\mu A_\nu -\partial_\nu A_\mu + g[A_\mu
,A_\nu]$, to its dual two-form,
\begin{equation}
  \label{SDcon} F_{\mu\nu} = \frac{1}{2}
  \varepsilon_{\mu\nu\rho\sigma}F^{\rho\sigma} \,.
\end{equation}
Solutions to the self-dual condition automatically satisfy the
Yang-Mills equation of motion due to the Bianchi identity,
\begin{equation}
D^\mu F^{\mu\nu} = \frac{1}{2} \varepsilon_{\mu\nu\rho\sigma}D^\mu F^{\rho\sigma} = 0 \,.
\end{equation}
Furthermore, field configurations that satisfy \cref{SDcon} correspond to instantons which are probed by the path integral in the full non-perturbative description of Yang-Mills. To construct a manifestly perturbative action, one begins by
transforming from Cartesian coordinates to light cone coordinates,
$(t,x,y,z) \rightarrow (u,v,w,\bar{w})$, via
\begin{equation}
u = t+x,\quad v=t-x, \quad w = y+iz, \quad \bar{w} = y-iz\, ,
\end{equation}
which yields the invariant line element,
\begin{equation}
ds^2 = 2(dudv - dwd\bar{w}) = dt^2 - d\vec{x}\cdot d\vec{x} \,.
\end{equation}
Applying this coordinate transformation to the self-dual condition of
\cref{SDcon}, produces three constraint equations on the field strength,
\begin{align} \label{SDLightCone1}
F_{uw} &=0
\\
 \label{SDLightCone2}
 F_{uv} &= F_{w\bar{w}}
\\
 \label{SDLightCone3}
 F_{v\bar{w}} &= 0\,.
\end{align}
In light-cone gauge, $A_u=0$, we the self-duality condition restricts the gauge field as follows:
\begin{align}
  A_u=0 \text{ and } (\ref{SDLightCone1}) \quad&\Rightarrow \quad A_w=0\\
  A_w=0\,,\,A_u=0 \text{ and } (\ref{SDLightCone2}) \quad&\Rightarrow \quad \partial_u A_v = \partial_w A_{\bar{w}} \label{eq:sd-divs}\,.
\end{align}
The last constraint is satisfied by writing the gauge field in
terms of a single scalar field $\Psi$ via
\begin{equation}
  \label{SDDef}
A_v = \frac{1}{2} \partial_w \Psi ~~~ \text{ and } ~~~ A_{\bar{w}} = \frac{1}{2} \partial_u \Psi \,.
\end{equation}
Using this definition on the first self-dual constraint equation
yields the following equation of motion for the dynamical scalar
field:
\begin{equation}
  \label{SDEOM}
(\ref{SDLightCone1}) \text{ and } (\ref{SDDef}) \quad  \Rightarrow\quad \Box \Psi + ig [\partial_u \Psi, \partial_w \Psi ]=0 \,.
\end{equation}
Adding back in the anti-holomorphic field $\bar{\Psi}$ as a Lagrange
multiplier, we obtain the following Lagrangian for
self-dual Yang-Mills theory:
\begin{equation}
  \mathcal{L}^{\text{SDYM}} = (\partial \bar{\Psi})(\partial \Psi) -i g \bar{\Psi} [\partial_u \Psi, \partial_w \Psi ] \,.
\end{equation}
The color-ordered cubic Feynman rule for this theory can be
immediately read off as
\begin{equation}
\cubic{hgrey0}{fermion2}{}{fermion2}{}{fermion2}{} =X(k_1,k_2) \equiv k_{u,1}k_{w,2} - k_{w,1}k_{u,2} \,.
\end{equation}
In this form, the Lagrangian and Feynman rules are manifestly
color-dual \textit{off-shell}.
The effect of \cref{SDcon} on the full YM Lagrangian is to decouple the anti-MHV three-point vertex from the theory, leaving only the MHV three-point vertex.
Before proceeding, it is important to
note that the mass-dimension of the theory appears to differ from that
of Yang-Mills -- we comment on this in detail in \cref{Conventions}.

\subsection{YZ-theory}
Self-dual Yang-Mills involves explicit reliance on the spacetime
dimension, which makes the theory poorly suited for constructing
dimensionally-regulated loop-level amplitudes.  On the other hand,
$YZ$-theory of \cite{Cheung:2016prv} is an honest $D$-dimensional theory that captures the
classical physics of NLSM while
manifesting the duality between color and kinematics.

One can understand $YZ$-theory as a particular dimensional reduction of Yang-Mills in $D=2d+1$ dimensions \cite{Cheung:2017yef}, down to $d$-dimensions. 
Starting with the Yang-Mills Lagrangian of \cref{eq:ym-lag}, one can redefine the gauge fields, $A_M$, in terms of $X$, $Y$ and $Z$ fields:
\begin{align}
X_M &= (X_\mu,0,-iX_\mu) 
\\
Y_M &= (0,Y,0) 
\\
Z_M &= (Z_\mu,0,iZ_\mu) 
\end{align}
where $X$ and $Z$ are $d$-vectors and $Y$ is a scalar field. Considering the conjugate nature of the $XZ$ propagator, in this work we make the replacement $X\rightarrow \bar{Z}$. Plugging in this redefinition of the gauge fields produces the following Lagrangian up to cubic order in the interactions
\begin{equation}
\mathcal{L}^{\text{YZ}} =\frac{1}{2} (\partial Y)^2 + (\partial Z)(\partial \bar{Z}) - g f^{abc} \left( \bar{Z}_{\mu\nu}Z^{\mu} Z^\nu + [Y,\partial_\mu Y] Z^\mu \right)\, .
\end{equation}
The Feynman rules for this Lagrangian are simply
\begin{equation}
\Acubic{hgrey0}{}{nhpRed}{}{nhpRed}{fermion2}{} = i (\varepsilon_3 p_2)\qquad \cubic{hgrey0}{fermion2}{}{fermion2}{}{fermion2}{} =i(\varepsilon_1 p_{3})(\varepsilon_2\bar{\varepsilon}_3) - i(\varepsilon_2 p_{3})(\varepsilon_1\bar{\varepsilon}_3) \, .
\end{equation}
Note that the pure vector vertex on the right is \textit{exactly} what was obtain from semi-abelian YM in \cref{eq:semiYM}. We now review of the construction of NLSM numerators at tree-level from $YZ$-theory. In this construction, we can define $D$-dimensional generators of the kinematic algebra as follows,
\be\label{eq:FeynmanRuleYYZ}
T^a_{ij}= i \varepsilon_a(p_i-p_j)\,,
\end{equation}
where momentum conservation requires,
\begin{equation}
p_a + p_i + p_j =0\, .
\end{equation}
The kinematic half-ladder diagrams then take on the following concise form
\begin{equation}
n^{\text{NLSM}}_{(i|a_1a_2...a_n|j)} = {}_i\langle T^{a_1}T^{a_2}\cdots T^{a_n}\rangle_j\, .
\end{equation}
Since there are no pole cancelling factors of $s_{ij} = (p_i+p_j)^2$, this definition of the kinematic algebra is manifestly cubic. Thus, the kinematic structure constants defined in terms of these generators are invariant under generalized gauge freedom because there are no quartic vertices. They can be defined implicitly through
\begin{equation}
[T^a,T^b]_{ij}= F^{a}_{\,b|c}T^c_{ij}\, .
\end{equation}
Given this definition, the Feynman rule associated with kinematic structure constant is
\be\label{eq:FeynmanRuleXZZ}
i F^{a}_{\,b|c} = (\varepsilon_b p_{ab})(\varepsilon_a\bar{\varepsilon}_c) - (\varepsilon_a p_{ab})(\varepsilon_b\bar{\varepsilon}_c)\, ,
\end{equation}
where $p_{ab}=p_a+p_b$ and $\varepsilon$ and $\bar{\varepsilon}$ are the polarizations of the $Z$-vectors particle and its conjugate field, respectively. The gauge-fixed state sum for this theory is simply
\begin{equation}
\sum_{\text{states}} \varepsilon^{\,\mu}_{(p)}\bar{\varepsilon}^{\,\nu}_{(-p)} = \eta^{\mu\nu}\, .
\end{equation}
Notice that the vector state sum is gauge fixed since the $YZ$ model explicitly chooses Lorenz gauge for the $Z$ particles, $\partial_\mu Z^\mu=0$.
Tree-level NLSM amplitudes are recovered from the kinematic structure constants by plugging in
\be\label{eq:onShellZStates}
\varepsilon^\mu_{(p)} = p^\mu \qquad \bar{\varepsilon}^\mu_{(p)} = \frac{q^\mu}{pq}
\end{equation}
 for the on-shell polarizations for $Z$ and $\bar{Z}$, respectively, where $q^2=0$ is some null reference momentum.
 The tree-level pion amplitude can then be defined in two equivalent ways,
\begin{equation}
\label{eq:TwoNLSMMethods}
A^{\text{NLSM}} = A(...,Y,...,Y,...) ~~~ \text{ and } ~~~ A^{\text{NLSM}}= A(...,\bar{Z},...) \, ,
\end{equation}
where the ellipses denote additional on-shell $Z$-particles. In a suitable gauge, the kinematic numerators in the latter definition for pion scattering are equivalent to those of $J$-theory, where $\bar{Z}$ corresponds to the root leg of $J$-theory \cite{Cheung:2021zvb}.

It is instructive to see how both of the constructions in \cref{eq:TwoNLSMMethods} produce valid tree-level amplitudes for the pion. First we will start with $Y$ particles on legs 1 and 4. Applying the Feynman rules above, and plugging in on-shell states for the $Z$-particles produces the following $s$- and $t$-channel numerators:
\begin{align}
n^{YY}_s &= (T^2T^3)_{14} = s_{12}^2 
\\
 n^{YY}_t &=  F^{3}_{\,2|X}T^X_{14}  = s_{14}(s_{13}-s_{12})\, .
\end{align}
Plugging these numerators into the ordered amplitudes $A(s,t)$ yields the desired result,
\be\label{eq:NLSMYZ4point}
A^{YY}_{(s,t)} = \frac{n^{YY}_s}{s_{12}}+\frac{ n^{YY}_t }{s_{14}} = s_{13}\, .
\end{equation}
For the $Z$ and $\bar{Z}$ configuration, the numerators are 
\begin{align}
n^{\bar{Z}Z}_s &= {}_4\langle F^{3}F^{2}\rangle_{1} =  s_{12}(s_{14}-s_{13})p_2^{\mu_1}+s_{12}^2(p_3-p_4)^{\mu_1}
\\
 n^{\bar{Z}Z}_t &=   {}_2\langle F^{3}F^{4}\rangle_{1}  = s_{14}(s_{12}-s_{13})p_4^{\mu_1}+s_{14}^2(p_3-p_2)^{\mu_1}\, ,
\end{align}
where we have used the shorthand
\begin{equation}
{}_x \langle F^{a_1}F^{a_2}\cdots F^{a_n}\rangle _{y} \equiv F^{a_1}_{\,x|b_2}F^{a_2}_{\,b_2|b_3}\cdots F^{a_n}_{\,b_n|y}\, .
\end{equation}
The polarization vector of the $\bar{Z}$ particle has not been contracted into the numerators above, resulting in the free $\mu_1$ index.
These numerators produce the partial amplitude
\begin{equation}
A^{\bar{Z}Z}_{(s,t)} = \frac{n^{\bar{Z}Z}_s}{s_{12}}+\frac{ n^{\bar{Z}Z}_t }{s_{14}} = -s_{13}(p_2+p_3+p_4)^{\mu_1} = s_{13} \,p_1^{\mu_1}\, .
\end{equation}
Plugging in the on-shell polarization of the conjugate field in \cref{eq:onShellZStates} produces precisely the desired result of \cref{eq:NLSMYZ4point}. Indeed, this construction is valid to all multiplicity at tree-level. One can see this by considering the only two possible factorization channels that contribute to each of these amplitudes, the $YY$ cut and the $\bar{Z}Z$ cut,
\begin{align} \label{YZCut1}
A(...,Y,...,Y,...) &\rightarrow A(...,Y,...,Y)A(Y,...,Y,...)
\\
&\rightarrow A(...,Y,...,Y,...,Z)A(\bar{Z},...) \, . \label{YZCut2}
\end{align}
Both factorization channels are valid descriptions of NLSM amplitudes when plugging in the $Z$ and $\bar{Z}$ on-shell states. As long as we plug in valid on-shell states for $Z$ and $\bar{Z}$ from \cref{eq:onShellZStates}, these numerators will produce NLSM ordered amplitudes. Given this special property of $YZ$-theory (namely that the $Z\bar{Z}$ and $YY$ constructions are equivalent), the $Y$ particle need not be considered when constructing NLSM amplitudes on shell. This is essentially what $J$-theory does, as we will show below.

\subsection{$J$-theory}
$J$-theory is obtained through a first order reformulation of NLSM in terms of the chiral current \cite{Cheung:2021zvb}.\footnote{For related formulations of NLSM that do not
treat color-kinematics duality see \cite{Freedman:1980us,
  Slavnov:1971mz}.} Two constraints are placed on the chiral current.  First, the field strength associated with $J_\mu$ should vanish,
\begin{equation}
\label{eq:FieldStrengthJ}
F_{\mu\nu}(J) = \partial_\mu J_\nu - \partial_\nu J_\mu + g[J_\mu , J_\nu]=0\, ,
\end{equation}
which then implies that $J^\mu$ is pure gauge $J_\mu = U \partial_\mu U^{-1}$.
The second condition is that the chiral current should be in Lorenz gauge,
\begin{equation}
\label{eq:LorenzGaugeJ}
\partial_\mu J^\mu=0\, ,
\end{equation}
which then yields the NLSM equation of motion $\partial_\mu (U \partial^\mu U^{-1})=0$.
Taking a linear combination of \cref{eq:FieldStrengthJ} and \cref{eq:LorenzGaugeJ} yields
\begin{equation}
\label{eq:JTheoryEOM}
\Box J^\mu +g f^{abc} J^\nu \partial_\nu J^\mu = 0.
\end{equation}
Integrating in an auxiliary field $\bar{J}^\mu$ as a Lagrange multiplier trivially produces the Lagrangian of semi-abelian YM \cref{eq:semiYM} but with the gauge fields renamed to $J_\mu$ and $\bar{J}_\mu$.
Since both theories are in Lorenz gauge this means that \emph{$J$-theory and semi-abelian YM are identical}. Furthermore, this theory is one-to-one with the amplitudes of $YZ$-theory when only considering on-shell $Z$$\bar{Z}$ states. As such, the single vertex in this theory is the same as the pure vector vertex of $YZ$ theory in Lorenz gauge,
\begin{equation}
\cubic{hgrey0}{fermion2}{}{fermion2}{}{fermion2}{} =i(\varepsilon_1 p_{2})(\varepsilon_2\bar{\varepsilon}_3) - i(\varepsilon_2 p_{1})(\varepsilon_1\bar{\varepsilon}_3)\, .
\end{equation}
By calculating an off-shell four-point correlation function it is possible to check that the Lagrangians of $J$-theory and semi-YM theory automatically satisfies color-kinematics duality, as described in \cref{offShellCK}.
As might be expected, the kinematic algebra is the diffeomorphism algebra \cite{Cheung:2021zvb}. Note that the different external particles break manifest Bose symmetry.

\subsection{Zakharov-Mikhailov theory}
\label{sec:ZMTheory}
While $J$-theory manifests color-kinematics duality, it only does so at
tree level because $J$-theory breaks manifest Bose symmetry.  Manifest
color-kinematics duality can be achieved at all loop orders by
restricting $J$-theory to two spacetime dimensions where the chiral
current can be dualized to a scalar
$J^\mu = \varepsilon^{\mu\nu} \partial_\nu \varphi$, which trivially
enforces Lorenz gauge.  After plugging $\varphi$ into
\cref{eq:JTheoryEOM} and rearranging, another copy of $\varphi$
(rather than some new field) can be integrated in to obtain the
Zakharov-Mikhailov (ZM) Lagrangian,
\begin{equation}
\label{eq:ZMLagrangian}
\mathcal{L}^{\text{ZM}} = \frac{1}{2}(\partial \varphi)^2 + g f^{abc} \varphi^a \varepsilon^{\mu\nu}(\partial_\mu \varphi^b)( \partial_\nu \varphi^c) .
\end{equation}
This procedure makes it clear that ZM and NLSM encode the same
classical physics, since their equations of motion are dual to each
other.  Furthermore, ZM theory can also be motivated from SDYM theory.
The power counting of the interaction term in the SDYM equation of
motion \cref{SDEOM} makes it clear that the theory is critical in 2D.
This suggests that the two lightcone coordinates $u$ and $w$ should be
identified as the coordinates of the 2D space.  The 4D d'Alembertian
in lightcone coordinates,
$\Box = 4 (\partial_u \partial_v - \partial_w \partial_{\bar{w}})$,
cannot be interpreted as the 2D equivalent, so the kinetic term of
SDYM must be replaced with the correct 2D version. However, changing the kinetic term in \cref{cubicCKLag} will have no effect on the Jacobi identity of \cref{jacID}, and thus does not interfere with color-kinematics duality. 

The color-stripped Feynman rule for the three-point vertex can be read
off from the Lagrangian \cref{eq:ZMLagrangian}, giving
\begin{equation}
\langle ab\rangle \equiv p_a^\mu \varepsilon_{\mu\nu} p_b^\nu\, .
\end{equation}
Unlike $J$-theory and $YZ$-theory, the color-stripped vertex is functionally antisymmetric on all legs
\begin{equation}
\cubic{hgrey0}{}{nhpBlue}{}{nhpBlue}{}{nhpBlue} = \langle 12\rangle = \langle 23\rangle = \langle 31\rangle = -\langle 21\rangle \, .
\end{equation}
This key property is due to momentum conservation alone.
To prove off-shell color kinematics duality for any multiplicity and
loop order it is enough to observe that the sum of the off-shell $s$-,
$t$-, and $u$-channel numerators sum to zero
\begin{equation}
\label{eq:ZMJacobi}
\langle ab \rangle \langle cd\rangle +\langle ac \rangle \langle db\rangle +\langle ad \rangle \langle bc\rangle =0 
\end{equation}
by virtue of the 2D Levi-Civita identity
$\varepsilon_{\mu\nu}\varepsilon_{\rho\sigma} =
\eta_{\mu\rho}\eta_{\nu\sigma}-\eta_{\mu\sigma}\eta_{\nu\rho}$.
Amusingly, \cref{eq:ZMJacobi} is simply a manifestation of the
Schouten identity.
Since ZM theory manifests color-kinematics duality without any
reference to the on-shell condition, a mass term can be added to ZM
theory without spoiling color-kinematics duality \cite{Cheung:2022mix}. In general, if the
on-shell conditions are not used, color-kinematics duality is
unaffected by the details of the propagator structure.  

Of course, dramatically altering the pole structure could spoil locality in the
double-copied theory.  As already mentioned, the only difference
classically between ZM and SDYM is the propagator structure.  While
both theories are color dual for the same reason, altering the
propagators has profound consequences since SDYM tree amplitudes
vanish and manifest color-kinematics does not persist to all loop
orders.  Color dual theories with the same interactions but different
kinetic terms have appeared in the literature before.  For example,
$J$-theory and the theory of non-Abelian fluids presented in
\cite{Cheung:2020djz} differ only in their propagators. Specifically,
taking the static limit of the fluid's $(D+1)$-dimensional kinetic
term $\partial_t - \nabla^2$ and Wick rotating one of the coordinates
yields the relativistic d'Alembertian of $D$-dimensional $J$-theory.

While the color-dual nature of ZM theory is rather elegant, the
loop-level construction introduces complications when applying
dimensional regularization. Similar to the difficulties of
renormalizing chiral fermions \cite{tHooft:1972tcz}, there is ambiguity
in promoting the integrands to formal $D$-dimensional expressions. As
such, in our two-loop bootstrap of NLSM we only used the ZM
integrands as a mechanism for reducing the residual gauge freedom in
our final solution.

\subsection{Chern-Simons theory}
\label{sec:CSTheory}
The final theory that is unified -- through one loop -- by the
semi-abelian Yang-Mills construction is Chern-Simons (CS) theory. As
we will show below, plugging in appropriate on-shell states to the
semi-YM numerators that we constructed in the text will produce
precisely the kinematic numerators for CS theory. Much like SDYM and
ZM theory, the CS action has a preferred integer dimension of $D=3$,
\begin{equation} S_{\text{CS}} = \int
d^3 x \,\text{tr}\left[\epsilon^{\mu\nu\rho} \left(A_\mu \partial_\nu
    A_\rho + \frac{2}{3}A_{\mu}A_{\nu}A_{\rho}\right)\right]
\label{eq:CSAction}
\end{equation}
where $\epsilon^{\mu\nu\rho}$ is a 3D Levi-Civita symbol. The
color-dual properties of this theory were previously studied in
Ref.~\cite{Ben-Shahar:2021zww}, which demonstrated that
color-kinematics duality is manifest off-shell.  At the classical
level, CS is clearly related to $J$-theory (and thus semi-YM) since
the CS equation of motion, \be \epsilon^{\rho \mu \nu}F_{\mu\nu} =0
\Leftrightarrow F_{\mu\nu}=0 \, , \ee and the Lorenz gauge choice, \be
\partial_\mu A^\mu=0 \, , \ee are the same as the $J$-theory
conditions in \cref{eq:FieldStrengthJ,eq:LorenzGaugeJ}, respectively.

The correspondence between CS and semi-YM can be made more precise at the level of the Feynman rules.
The color-stripped three-point vertex and propagator can be read off from the CS action in \cref{eq:CSAction} producing
\begin{equation}
  \label{CSFrule}
\cubic{hgrey0}{nhpRed}{}{nhpRed}{}{nhpRed}{} = \epsilon^{\mu\nu\rho} \varepsilon^{\mu}_1\varepsilon^{\nu}_2\varepsilon^{\rho}_3 \qquad \qquad \propFrule{$A^\nu$}{$A^\mu$}{$k \rightarrow $}{nhpRed} = \frac{i}{k^2} \epsilon^{\mu\nu\rho} k^\rho \, .
\end{equation}
At first glance, this seems far removed from the semi-abelian
Yang-Mills analogs in \cref{semiYMvert} and \cref{semiYMprop}. After
all, semi-YM makes no reference to an antisymmetric 3-tensor and the
mass dimensions of the vertices do not agree. However, to construct
the full on-shell amplitudes of CS theory, we must plug in the Fourier
transforms of the on-shell currents, $\bar{\varepsilon}(p)$, as
described in Ref.~\cite{Ben-Shahar:2021zww}.  These are related to the
polarizations, $\varepsilon_i$, of the Feynman rules above as follows:
\begin{equation}
\varepsilon^{\mu}(k) = \epsilon^{\mu\nu\rho} k^\nu \bar{\varepsilon}^\rho(k)\, .
\end{equation}
With this definition of the on-shell states, it is clear how the Feynman rules of \cref{CSFrule} are related to those of \cref{semiYMvert} and \cref{semiYMprop} -- we must simply canonicalize the CS kinetic term by absorbing $ \epsilon^{\mu\nu\rho} k^\rho$ into the definition of the three-point vertex:
\begin{equation}
 \propFrule{$A^\nu$}{$A^\mu$}{$k \rightarrow $}{nhpRed} = \frac{i}{k^2} \epsilon^{\mu\nu\rho} k^\rho
\quad \longrightarrow \quad  \propFrule{$A^\nu$}{$\bar{A}^\mu$}{$k \rightarrow $}{fermion2} = \frac{i}{k^2} \eta^{\mu\nu}\,.
\end{equation}
This has the effect of orienting the three-point vertex of \cref{CSFrule} by making the replacement $\varepsilon_i \rightarrow \epsilon^{\mu\nu\rho} k_i^\nu \bar{\varepsilon}^\rho_i$ on one of the legs,
\begin{align}
  \cubic{hgrey0}{nhpRed}{}{nhpRed}{}{nhpRed}{} \rightarrow \cubic{hgrey0}{}{fermion2}{}{fermion2}{}{fermion2} =
  \epsilon^{\mu\nu\rho} \varepsilon^{\mu}_1\varepsilon^{\nu}_2 (\epsilon^{\rho \alpha \beta} k_3^\alpha \bar{\varepsilon}^\beta_3) =    (\varepsilon_1\bar{\varepsilon}_3)(\varepsilon_2 k_3)
  -  (\varepsilon_2\bar{\varepsilon}_3)(\varepsilon_1 k_3)\, ,
\end{align}
where we have used the identity
$\epsilon^{\mu\nu\rho}\epsilon^{\rho\alpha\beta} = \eta^{\mu \alpha}
\eta^{\nu \beta} - \eta^{\mu \beta} \eta^{\nu \alpha}$. As desired,
the redefined CS vertex is precisely the three-point vertex of semi-YM
in \cref{semiYMvert}. Clearly the $D$-dimensional amplitudes of
semi-YM can be mapped to the 3D observables of CS theory by making the
following replacement on all of the external states of semi-abelian
YM: \be \varepsilon^\mu_i \rightarrow \epsilon^{\mu\nu\rho} k^\nu_i
\bar{\varepsilon}^\rho_i \qquad\qquad \bar{\varepsilon}_i^\mu
\rightarrow \bar{\varepsilon}^\mu_i \, .  \ee Upon making this
replacement, the amplitudes of $D$-dimensional semi-YM are mapped onto
CS theory in $D=3$,
$\mathcal{A}^{\text{semi-YM}}\rightarrow
\mathcal{A}^{\text{CS}}$. While this may seem like a marginal gain at
tree-level, by using our $D$-dimensional construction at one-loop, the
manifestly color-dual integrands of \cref{SDYMnGon} can resolve
regularization ambiguity of CS theory due to its explicit dependence
on $D=3$. For more background on CS theory, we refer the reader to
Ref.~\cite{Ben-Shahar:2021zww}.

\section{Spinor-helicity and conventions}\label{Conventions}

We use the same conventions as Ref.~\cite{jjmcTASI2014}, which
we quote now. The component-wise definitions of the spinor brackets are
\begin{align}
\langle ab \rangle &= \frac{(a_1 + i a_2)(b_0+b_3)-(b_1 + i b_2)(a_0+a_3)}{\sqrt{(a_0+a_3)(b_0+b_3)}}\,,
\\
[ab] &= \frac{(b_1 - i b_2)(a_0+a_3)-(a_1 - i a_2)(b_0+b_3)}{\sqrt{(a_0+a_3)(b_0+b_3)}}\,,
\end{align}
where the $a_i$ are components of the four-vector,
$k^\mu_a = (a_0,a_1,a_2,a_3)$. For massless momenta, $k_a$ and $k_b$,
one can apply the on-shell condition, $k_a^2=k_b^2=0$, to show that
\begin{equation}
s_{ab} = (k_a+k_b)^2= \langle ab \rangle[ba]\,.
\end{equation}
Four-dimensional polarization dot products with fixed helicity states
are mapped to
\begin{equation}\label{eq:4DPols}
\begin{aligned}
k_a \cdot \varepsilon_b^{(+)} &= \frac{\langle q a \rangle[ab]}{\sqrt{2}\langle q b\rangle}\,,
\qquad\quad \qquad
k_a \cdot \varepsilon_b^{(-)} = -\frac{[qa]\langle ab\rangle}{\sqrt{2}[qb]}\,,
\\
\varepsilon_a^{(-)}\cdot \varepsilon_b^{(+)} &= - \frac{\langle q a\rangle [qb]}{ [qa]\langle q b\rangle} \,,
\qquad \qquad
\varepsilon_a^{(\pm)}\cdot \varepsilon_b^{(\pm)} = 0 \,,
\end{aligned}
\end{equation}
Note that given the above definition, the spinor helicity variables
carry the same mass dimension for both the angle and square
brackets. In terms of the light-cone coordinates that we employed in
\cref{sec:CKLagrangians}, the angles and squares can be rewritten as
\begin{align}
\langle ab \rangle &= \frac{a_w b_u-a_u b_w}{\sqrt{a_ub_u}}\,,
\\
[ab] &= \frac{b_{\bar{w}}a_u-a_{\bar{w}}b_u}{\sqrt{a_ub_u}}\,.
\end{align}
The conventions we use in the text where the holomorphic spinor,
$X(p,k)$, carries extra mass dimension amounts to shifting
the bracket by an overall factor,
\begin{align}
\langle ab \rangle &\rightarrow a_w b_u-a_u b_w \equiv X(a,b)\,,
\\
[ab] &\rightarrow \frac{b_{\bar{w}}a_u-a_{\bar{w}}b_u}{a_ub_u}\equiv Q(a,b)\, .
\end{align}
With this choice we still obtain the same completeness relation for the SDYM spinors,
\begin{equation}
X(a,b)Q(b,a) = s_{ab}\, .
\end{equation}

\section{Regulating BEL integrals}\label{BELreg}
This appendix covers the bubble-on-external-leg (BEL) integrals that result from the $n$-gon numerator of $Y\!Z$ theory.
Before covering the BEL integrals themselves, we will need the appropriate numerator.
Weight counting tells us that the $n$-gon master
numerator must have $n$ on-shell $Z$-particles. Unitarity requires
that there are three distinct contributions to the $n$-gon: one from an off-shell $Y$-loop particle and then two more from
different orientations of a $\bar{Z}Z$-loop.  Thus, $YZ$ theory gives
us the following one-loop $n$-gon numerator:
\begin{equation}
N^{n\text{-gon},YZ}_{(12...n)} = \langle T^{1}T^{2}\cdots T^{n}\rangle+ \langle F^1F^2\cdots F^n\rangle
\end{equation}
where $\langle\,\cdots\rangle$ indicates an internal contraction over the $YY$ and
$\bar{Z}Z$ loops and $T$ and $F$ were given in \cref{eq:FeynmanRuleYYZ} and \cref{eq:FeynmanRuleXZZ}, respectively.
In terms of the external momenta $k_i$ and the loop momenta $\ell_i$, where $\ell_i$ flows into $k_i$ and out of $k_{i-1}$, the NLSM numerator is
\begin{equation}
N_{(12...n)}^{\text{NLSM}}=\langle T^{1}T^{2}\cdots T^{n}\rangle =2^n (\ell_1 k_1)(\ell_2 k_2) \cdots (\ell_n k_n)
\end{equation}
which is the same as \cref{NLSMnGon} after using $2(k_i \cdot \ell_i) = \ell_{i}^2-\ell^2_{i-1}$. Similarly, there is a pure vector contribution that is identical to the semi-abelian YM $n$-gon numerator that we constructed in \cref{SDYMnGon} after projecting external states along longitudinal modes $\varepsilon \rightarrow k$,
\begin{equation}
 \langle F^1F^2\cdots F^n \rangle= D (\ell_1 k_1)(\ell_2 k_2) \cdots (\ell_n k_n) + \mathcal{O}(D^0)\, .
\end{equation}
The dimension-dependent factor essentially counts that number of
internal vector states. While this $n$-gon is manifestly color-dual,
it does not produce the right cuts for NLSM. However, the scalar
contribution, that comes dressed with an overall factor $D$
\textit{does} manifest the duality globally, and satisfies all the
desired pion cuts due to the factorization of \cref{YZCut1} and \cref{YZCut2}. In order for the Feynman rules for $Y\!Z$ theory
compute one-loop color-dual numerators consistent with NLSM cuts, one would have to add additional states to cancel off the spurious poles, while
preserving color-kinematics duality. We leave this as a direction of
future work.

While the $\bar{Z}Z$ vector loop spoils color-kinematics off-shell, the $\bar{Y}Y$ loop alone gives us the desired expression for the $n$-gon. However, as we noted in the text, the $n$-gon numerator has the strange property that it produces \textit{non-vanishing} bubble-on-external-leg (BEL) graphs. However, the BEL diagram integrates to zero after integral reduction.
As an example, the four-point BEL diagram can be reconstructed from the $n$-gon,
\begin{equation}
N^{\text{BEL}}_{1|2,34} = N^{\text{NLSM}}_{(1234|\ell)}-N^{\text{NLSM}}_{(1243|\ell)}-N^{\text{NLSM}}_{(1342|\ell)}+N^{\text{NLSM}}_{(1432|\ell)}\, ,
\end{equation}
where we define the loop momentum to be in between the left most and right most leg on the box. Plugging in particular values for $\ell_i$, the BEL reduces to
\begin{equation}
N^{\text{BEL}}_{1|2,34} = s_{12} (\ell+k_1)^2 \ell^{\mu} k_1^{\nu} k_2^{[\mu} k_{[34]}^{\nu]} \,.
\end{equation}
Notice there is an overall factor that cancels one of the propagators. Plugging this numerator into the amplitude produces an integral of the form
\begin{equation}
\mathcal{I}^{\text{BEL}}_{1|2,34} = s_{12} k_1^{\nu} k_2^{[\mu} k_{[34]}^{\nu]} \int \frac{d^D \ell}{i\pi^{D/2}} \frac{\ell^\mu }{\ell^2-\mu^2} \sim   s_{12}(s_{13}-s_{14}) (\mu^2)^{D/2}\, ,
\end{equation}
where we have introduced a mass regulator that will be proportional to the on-shell momentum inside the BEL, $\mu^2 \equiv k_1^2$. Thus, in sufficiently large dimension, $D>2$, this integral suppresses the $\mu^{-2}$ divergence appearing in the denominator of the BEL diagram. 

\bibliographystyle{JHEP}
\bibliography{Refs_2loopNLSM}
\end{document}